\RequirePackage{fix-cm}
\documentclass[smallcondensed
]{svjour3}     
\usepackage{graphicx}

\usepackage{tikz}
\usetikzlibrary{patterns}
\usetikzlibrary{shapes.geometric, calc}
\usepackage[sort&compress]{natbib}

\journalname{}

\usepackage{amsmath,amssymb,amsfonts}
\usepackage{tcolorbox}

\usepackage{verbatim}
\usepackage{xspace}

\usepackage{etoolbox} 

\usepackage{thmtools}
\usepackage{mathtools}
\usepackage{thm-restate}
\usepackage{ulem}

\usepackage{tikz-cd}

\usepackage{hyperref}
\usepackage{xcolor}
\definecolor{darkgreen}{RGB}{0,100,0}

\definecolor{firebrick}{RGB}{178,34,34}
\definecolor{Andre_blue}{RGB}{0,120,220}
\definecolor{Andre_gray}{RGB}{180,200,200}

\usepackage{tikz} 
\usetikzlibrary{patterns}
\usetikzlibrary{shapes.geometric, calc}

\usepackage{amsthm}
\theoremstyle{plain}
\newtheorem{claim2}[theorem]{\textbf{Claim}}

\newtheorem{setting}[theorem]{\textbf{Setting}}

\theoremstyle{remark}

\DeclareMathOperator{\ax}{{ax}}

\newcommand{\R}{\mathbb{R}}
\newcommand{\N}{\mathbb{N}}

\newcommand{\M}{\mathcal{M}}
\newcommand{\Sphere}{\mathbb{S}}

\newcommand{\length}{\operatorname{length}}
\newcommand{\Su}{\mathcal{S}}
\newcommand{\Nor}{\operatorname{Nor}}
\newcommand{\Tan}{\operatorname{Tan}}
\newcommand{\rch}{\operatorname{rch}}
\newcommand{\ulog}{\operatorname{ulog}}
\newcommand{\conv}{\operatorname{conv}}

\newcommand{\Set}{\Su}
\newcommand{\CAT}{\operatorname{CAT}}

\title{Sets of reach $R$ are \texorpdfstring{$\operatorname{CAT}\left( 1/R^2\right)$}{CAT 1/R2}
}

\begin{document}
\author{Andr{\'e} Lieutier \and Mathijs Wintraecken}

\title{Sets of reach $R$ are \texorpdfstring{$\operatorname{CAT}\left( 1/R^2\right)$}{CAT 1/R2}
\thanks{
		Mathijs Wintraecken was supported by the Austrian science fund (FWF) grant No. M-3073, the ANR grant StratMesh (ANR-24-CE48-1899) and the welcome package from IDEX of the Universit{\'e} C{\^o}te d'Azur (ANR-15-IDEX-01). 
}
\thanks{The first part of this paper was presented at the Symposium on Computational Geometry 2026. }
}

\titlerunning{Sets of reach $R$ are \texorpdfstring{$\operatorname{CAT}\left( 1/R^2\right)$}{CAT 1/R2}} 

\institute{
\\
Andr{\'e} Lieutier \at No affiliation\\ \email{andre.lieutier@gmail.com}
\\
			Mathijs Wintraecken (corresponding author)\at 
			Inria centre d'Universit\'e C\^ote d'Azur \\
						 \email{mathijs.wintraecken@inria.fr}   
}


\maketitle


\begin{abstract} 
{
Positive reach underpins many results in computational geometry and topology. It is used for triangulation criteria, topological inference, and manifold learning. 
The geometric properties of these sets have therefore been studied intensely. In this paper we focus on the intrinsic metric properties of sets of positive reach. In particular we give an elementary proof of the fact that (closed) sets of reach at least $R$ are $\operatorname{CAT}(1/R^2)$. This fact was established earlier by Alexander and Bishop, using quite significant machinery. 
The fact that a space is $\operatorname{CAT}(1/R^2)$ offers many advantages, such as the uniqueness and continuity of sufficiently short geodesics, the existence of Fr{\'e}chet means and by extension the Rips based homotopy reconstruction of Hausmann is available. 
}
\end{abstract}

\maketitle

\section{Introduction}

Federer \cite{Federer} introduced the reach of a closed set $\Su \subset \mathbb{R}^d$ as the infimum of the distance from $\Su$ to the medial axis $\ax(\Su)$, i.e. the set of points in $\mathbb{R}^d$ for which the closest point in $\Su$ is not unique. Assumptions on the reach (and its local version the local feature size~\cite{Amenta1999}) underpin the correctness of many algorithms in computational geometry and topology. 
{
In this paper we consider the metric on sets of positive reach induced by shortest paths or geodesics distance. 
We provide an elementary proof of the fact that a set of at least reach $R$ is $\operatorname{CAT}(1/R^2)$. 

The first part of the proof is dedicated to the fact that geodesics in a set of positive reach are unique and continuous with respect to the end points as long as the geodesic is strictly shorter than $\pi R$. We note that this bound in tight, because attained for the sphere. The uniqueness and continuity of geodesics also follows from the fact that the space is $\operatorname{CAT}(1/R^2)$, but is required 
{as a first step} in the proof.

Unfortunately, when this paper was almost complete the authors discovered that the $\operatorname{CAT}(1/R^2)$ property was established by Alexander and Bishop in \cite{Alexander2006}. The language used by Alexander and Bishop is such that this result is not that easy to recognize as \cite[Theorem~1]{Alexander2006}, without looking at it quite closely. 
However, we believe that the different and elementary approach to the question in this paper may still be of interest. 

\subsection{Motivation} 
 

\paragraph*{Triangulation algorithms} $\phantom{3}$
The assumption of positive reach is central to almost all triangulation algorithms, see for example \cite{ChengDeyShewchuck, Dey, JDbook}, as well as manifold learning, see for example \cite{FeffermanFitting, fefferman2019GeometricWhitneyProblem, Eddie2018stability, sober2019manifold}. However these algorithms often consider manifolds of positive reach instead general sets of positive reach. If we want to extend our algorithm beyond the realm of manifolds to more general stratified spaces, starting with `well behaved' sets of positive reach\footnote{Sets of positive reach are stratified spaces~\cite{Federer}, where the stratification is understood in a weak sense. However, it is not clear that sets of positive reach are triangulable in general or additional assumptions are needed. This having been said, there are compact sets of (arbitrarily large) positive reach that are not (piecewise smoothly/linearly) homeomorphic to a finite simplicial complex, {which puts this outside of the realm of computation.}   
} or at least sets of which the strata have positive (local \cite{FullVerionc11}) reach understanding the geometry of those sets seems essential. 
Moreover, the existence of Fr{\'e}chet means in $\operatorname{CAT}(K)$ spaces makes it possible to define (possibly degenerate) intrinsic simplices, c.f. \cite{RiemSimp2015}.

\paragraph*{Homotopy reconstruction} $\phantom{3}$ 
The reconstruction of the homotopy of set of positive reach using \v{C}ech complexes has been studied extensively within computational topology since the appearance of the seminar work of Niyogi, Smale, and Weinberger \cite{niyogi2008}. Although originally formulated exclusively for smooth manifolds this has been extended to include first manifolds with boundary \cite{WangWang2020} and then general sets of positive reach\cite{attali2024tight}. 

Learning the homotopy type \cite{Hausmann, Latschev2001} of a (non-embedded) Riemannian manifold of bounded curvature or more generally a $\operatorname{CAT}(K)$ space, based on the Rips-complex (based on intrinsic distances), has attracted significant attention and has recently been quantified in \cite{komendarczyk2026topologicalstabilitylatschevtypereconstruction, Demystifying2024}, which rely on Fr{\'e}chet means. 
The quantified bounds on the curvature in terms of the reach allow for a more precise comparison between these different inference methods.  

\paragraph*{Metric learning} $\phantom{3}$
Another motivation comes from metric learning. In \cite{aamari2023optimal} this is explained as follows: 

``In the data analysis area, metric learning refers to the problem of finding a distance $d$ over the space of observations that is relevant for a given task at stake \cite{yang2006dmlSurvey,suarez2021tutorial}. For instance, in a supervised framework where one is provided with tuples of allegedly similar or dissimilar observations, the goal is to find a distance that is small on the similar tuples and large on the dissimilar ones. There is a wide range of existing methods in the literature, ranging from parametric (LSI \cite{xing2002distance}, MCML \cite{globerson2005mcml}, LDML \cite{guillaumin2009ldml} among others) to nonparametric (DMLMJ \cite{nguyen2017dmlmj}, kernel methods \cite{kwok2003idealizedKernels,chatpatanasiri2010kernelization}, to cite a few).

In an unsupervised setting, metric learning aims at finding a metric that takes into account the underlying geometry of the data. That is, it amounts to estimating the shortest-path (or geodesic) distance. Often, this is done via a dimension reduction technique: any low-dimensional embedding of the data gives rise to a new distance over the data in the embedded space. Existing algorithms include PCA, t-SNE \cite{hinton2002sne}, MDS \cite{cox2008mds}, Isomap \cite{tenenbaum2000isomap}, or MVU \cite{ariascastro2013mvu}. See \cite{suarez2021tutorial} for a thorough overview of the field.'' 

And indeed in \cite{aamari2023optimal} geodesics play a crucial role in establishing optimal bounds. {The stability of geodesics, which follows from the main result of this paper, could provide an essential tool in studying the stability of these methods. }

\subsection{Contribution} 
The main result of the paper is the following: 
\begin{restatable}{theorem}{MainCAT}\label{ThmCAT}  
If $\Set \subset \R^d$ is a closed set with reach at least $R>0$, then $\Set$ endowed with the 
 geodesic distance $d_{\Set}$ is a $\operatorname{CAT}\left( \frac{1}{R^2}\right)$ metric space.
\end{restatable}
The proof has two important stepping stones, which would in fact also follow from the main statement (once it is available). The first stepping stone is:
\begin{restatable}{theorem}{UniqueG}\label{prop:Unique}  
Let $\Su$ be a set of positive reach $R$, then for any two points $a,b\in \Su$ such that {the geodesic distance on $\Su$ satisfies} $d_\Su (a,b) <\pi R$ there is a unique minimizing geodesic in $\Su$ connecting $a$ and $b$. 
\end{restatable}
\noindent  \fbox{ \parbox{0.97\textwidth} { { We stress that in the context of sets of positive reach a minimizing geodesic is a curve of minimal length connecting the two points. In this paper we only consider minimizing geodesics and therefore will not explicitly repeat `minimizing' everywhere. More generally, a geodesic can be defined as a curve such that it is locally a minimizing geodesic (between the endpoints of a subsegment). In particular more than half of great circle on a sphere, is NOT a minimizing geodesic, although such a curve is a geodesic in the more general and Riemannian contexts.   }}}

\noindent The second stepping stone is a corollary of Theorem \ref{ThmCAT}  
\begin{restatable}{corollary}{ContinuousGCOne}\label{Cor:ContinuousGc1}  
Let $\Su$ be a set of positive reach $R$, then the geodesic connecting any two points $a,b\in \Su$ is continuous (in the $C^1$ sense) with respect to the endpoints provided that $d_\Su (a,b) <\pi R$. 
\end{restatable}
We stress that all results are tight, because the bounds are attained for the standard sphere of radius $R$ embedded in Euclidean space.

\subsection{Previous work} 
\paragraph*{$\textrm{CAT}(k)$ spaces} $\phantom{3}$
The terminology $\textrm{CAT}(k)$ was introduced by Gromov \cite{gromov1987hyperbolic} for metric spaces that satisfy some weak form of curvature bound. We'll now give a rough definition of $\textrm{CAT}(k)$ spaces, but refer to \cite[Section II.1]{bridson2013metric} for a precise introduction. 
Let $(\M, d_\M)$ be a geodesic metric space, that is a space where the length between two points is defined as the minimum of all lengths of continuous (rectifiable) curves connecting the two points. 

Such a minimizing curve is called a geodesic. 
A geodesic triangle in such a space consists of three points and the geodesics connecting those points. We can then consider a geodesic triangle with the same edge lengths in a (simply connected) space of constant curvature $k$ (that is a sphere, Euclidean space, or hyperbolic space). This triangle is called a comparison triangle. 
We say that the diameter of a simply connected space of constant curvature is $\pi / \sqrt{k}$ if $k>0$ and $\infty$ otherwise. We also refer to this as the diameter of the comparison space.  This geodesic triangle in $\M$ satisfies the $\textrm{CAT}(k)$ inequality if the distances between the points on the triangle are less than the corresponding distances between points on the comparison triangle in a space of constant curvature $k$. If this holds for any geodesic triangle of diameter less than the diameter of the comparison space, then we say that the entire space is $\textrm{CAT}(k)$. 

 %
 %
%

\paragraph*{Sets of positive reach as $\operatorname{CAT}(k)$ spaces} $\phantom{3}$
Lytchak \cite{lytchak2004geometry, lytchak2005almost} proved sets of positive reach are $\textrm{CAT}(k)$, for some $k$. 

As mentioned, this result was later quantified by Alexander and Bishop \cite{Alexander2006}, whose paper tells us,  albeit in a slightly different guise, that sets of reach $R$ are at least $\operatorname{CAT}(1/R^2)$. 
This paper reproves this result in a completely elementary fashion, 
in the sense that we only need some results by Federer \cite{Federer}, comparison theory of second order differential equations, see Berger \cite[Section 3.2]{Berger} for a very short and pedagogical discussion and proof, and Euler-Lagrange theory \cite{SpivakI}. 

\subsection{Future work} 

The fact that sets of reach $R$ are $\CAT(1/R^2)$ opens a significant toolbox that has been developed in mathematics over the last decades. There are some interesting questions within computational geometry and topology for which the $\CAT(1/R^2)$ is likely to help and that, as far as the authors are aware, are still open: 
\begin{itemize} 
\item We know that tangent cones of sets of positive reach are semi-continuous \cite{FullVerionc11} in the sense that there exists a neighbourhood such that the one-sided angle between the affine hulls of the tangent cones is bounded. We stress that for general sets of positive reach the one-sidedness of the angle is essential, because the dimension of the tangent cone can increase. 
We suggest to search for explicit Lipschitz 
bounds on the one-sided angle. {Here we need to explain what we mean by Lipschitz in this semi-continuous setting: We conjecture that for every $x \in \mathcal{S}$ there exists an open neighbourhood $U \subset \mathbb{R}^d$, such that for all $y \in U \cap \Su$ the one-sided angle from the affine hull of $\Tan(y ,\Su)$ to the affine hull of $\Tan(x, \Su)$ is bounded by $d_{\Su} (x,y) / \rch(\Su)$.  
 }

This question is motivated by stratification learning \cite{AdamStrata, bendich2010, bendich2007}.   
\item As mentioned in the motivation many learning algorithms are based on estimation of intrinsic distances. The $\CAT(1/R^2)$-property could help establish or improve stability guarantees for such algorithms. 
\end{itemize}

\section{Notation and preliminaries}

Although we often recall previous results and introduce notation when they are used, there are a couple of notations and results that are featured so often below that we recall them here.

We'll denote sets of positive reach by $\mathcal{S}$ and the closest point projection by $\pi_\Su$. We use the notation $B(c,r)$ for a ball in Euclidean spaces with centre~$c$ and radius~$r$. {For $a,b \in \Set$,  $d_{\Set}(a,b)$ denotes the geodesic distance between $a$ and $b$ in $\Set$.}
Curves are generally denoted by $\gamma$ and most importantly: 

\noindent  \fbox{ \parbox{0.95\textwidth} {We'll use Newton's notation for the derivatives of curves that is, $\dot{\gamma} = \frac{d}{dt} \gamma$ and $\ddot{\gamma} = \frac{d^2}{dt^2} \gamma$. } }

If we want to emphasize the end points ($a,b$) of a geodesic we write $\gamma_{a,b}$. 


{
Originally \cite{Federer}, the reach was defined as:
\[
\operatorname{rch}  (\Set) = \inf_{\sharp \pi_\Su (x) \geq 2} d(x, \Su ),
\]
where ``$\sharp$'' denotes the cardinality. }

The equivalent metric characterization of the reach of a set is the following: 
\begin{theorem}[Theorem 1 of \cite{TangentVarJournal}]\label{theorem:ReachEquivalentMetricDistorsion}
If $\Set \subset \R^d$ is a closed set, then
$$
\operatorname{rch}  (\Set) = \sup \left\{r > 0,  \, \forall a,b \in \Set, \, |a-b| < 2 r \Rightarrow d_{\Set}(a,b) \leq 2 r \arcsin \frac{|a-b|}{ 2 r} \right\},
$$
where the $\sup$ over the empty set is $0$ and $\operatorname{rch} (X)$ denotes the reach of $X$.
\end{theorem}


%

Theorem \ref{theorem:ReachEquivalentMetricDistorsion} implies the following: 
\begin{corollary}\label{lemma.geodesicExistPartA}

One has:

\begin{itemize} 
\item If {$R\leq \operatorname{rch}  (\Set)$ and} $a,b \in \Set$
 satisfy $|a-b| < 2 R$ 
then $d_{\Set}(a,b)< \pi R$.
\item Geodesics in a set $\Set$ of reach $R$, seen as curves in $\R^d$, have reach at least $R$.
\end{itemize} 
\end{corollary}
\begin{proof} 
The first statement follows because $\arcsin (1)= \pi/2$. The second statement is true because geodesics themselves satisfy the characterization. 
\end{proof} 

In \cite[Theorem 1]{TangentVarJournal} we have seen that manifolds of positive reach are $C^{1,1}$ with a $(1/R)$-Lipschitz derivative. 
Combining this with the previous result gives that:
\begin{corollary} \label{cor:TwiceDiffAlmostAnywhere}  
Geodesics in a set of reach $R$ are $C^{1,1}$ with a $(1/R)$-Lipschitz derivative and hence the derivative is twice differentiable almost everywhere. In particular, if $\gamma$ denotes a geodesic, we have $|\ddot{\gamma}(s)| \leq \frac{1}{R}$, whenever the second order derivative exists. 
\end{corollary}
\begin{proof}
As mentioned, the fact that geodesics are $C^{1,1}$ with a $(1/R)$-Lipschitz derivative follows from \cite[Theorem 1]{TangentVarJournal}.  
Thanks to the Rademacher theorem, see e.g. \cite[Theorem 3.1.6]{FedererBook}, the derivative is therefore twice differentiable almost everywhere.
\end{proof}


Next, we introduce the normal cone. We denote the scalar product in $\R^d$ by~$\langle .,. \rangle$.
\begin{definition}[Definitions 4.3 and 4.4 of \cite{Federer}] \label{def:4.3and4.4Fed} 
	If $\Su \subseteq \R^d$ and $p\in \Su$, then the generalized
	tangent space $\Tan(p,\Su)$
	is the set of all tangent vectors of $\Su$ at $p$. 
	It consists of all those $u \in \R^d$, such that either $u=0$ or for every $\varepsilon>0$ there exists a point $q \in  \Su$ with
	\begin{align}
		0<&\|q-p\|<\varepsilon &\textrm{and} & &\left\| \frac{q-p}{\|q-p\|}- \frac{u}{\|u\|} \right\| < \varepsilon.
		\nonumber 
	\end{align}
	The normal cone of $\Su$ at $p$ is the set 
	\begin{align*}
		\Nor(p,\Su)
	\end{align*}
	of all vectors $v \in \R^d$ such that  $\langle v,u \rangle \leq 0$ for all
		$u \in \Tan(p,\Su)$.
\end{definition}

\section{Uniqueness and continuity of geodesics} 
\subsection{Uniqueness} 
{The main idea of the proof is the following: Suppose that there are two geodesics $\gamma_1$ and $\gamma_2$ connecting two points is not unique then we can move the geodesic $\gamma_1$ a little in the direction $\gamma_2$, shortening the geodesic and thus yielding a contradiction. This perturbed geodesic, of course, needs to lie on the set of positive reach, which will be achieved by closest point projection onto the set of positive reach. We commence our proof with a couple of lemmas that will be instrumental in dealing with this projection.   }

We first recall a convexity result from \cite{TangentVarJournal}. {We recall that a set $C$ is geodesically convex if for any two points $c, c' \in C$ the geodesic connecting $c$ and $c$ is contained in $C$. }
\begin{corollary}[Corollary 1 of \cite{TangentVarJournal}]\label{corollary:LocalGeodesicConvexity}
Let $\Su \subset \R^d$ be a closed set with positive reach $R >0$.
Then, for any $r' < R$ and any $x \in \R^d$ 
then
$
\Su \cap B(x,r')
$
is geodesically convex in $\Su$.
\end{corollary}

\begin{lemma}\label{lemDistProjMidpoint}  
Let $\Su$ have reach $R$ and assume $p, q \in \Su$ with $|p-q| \leq 2R$. Define $x =  \lambda p + (1 -\lambda) q$. We have that 
\begin{align} 
d(x, \Su ) &= 
|x - \pi _\Su (x) | 
\nonumber 
\\
&\leq  R - \sqrt{ R^2 - \lambda (1-\lambda) |p-q|^2} 
\nonumber 
\\
& \leq \frac{\lambda (1-\lambda) |p-q|^2}{2R} + \mathcal{O}   \left(   \left(  \frac{\lambda (1-\lambda) |p-q|^2}{2R}\right)^2   \right) ,
\nonumber
\end{align} 
{with $d$ the Euclidean distance. }
\end{lemma} 
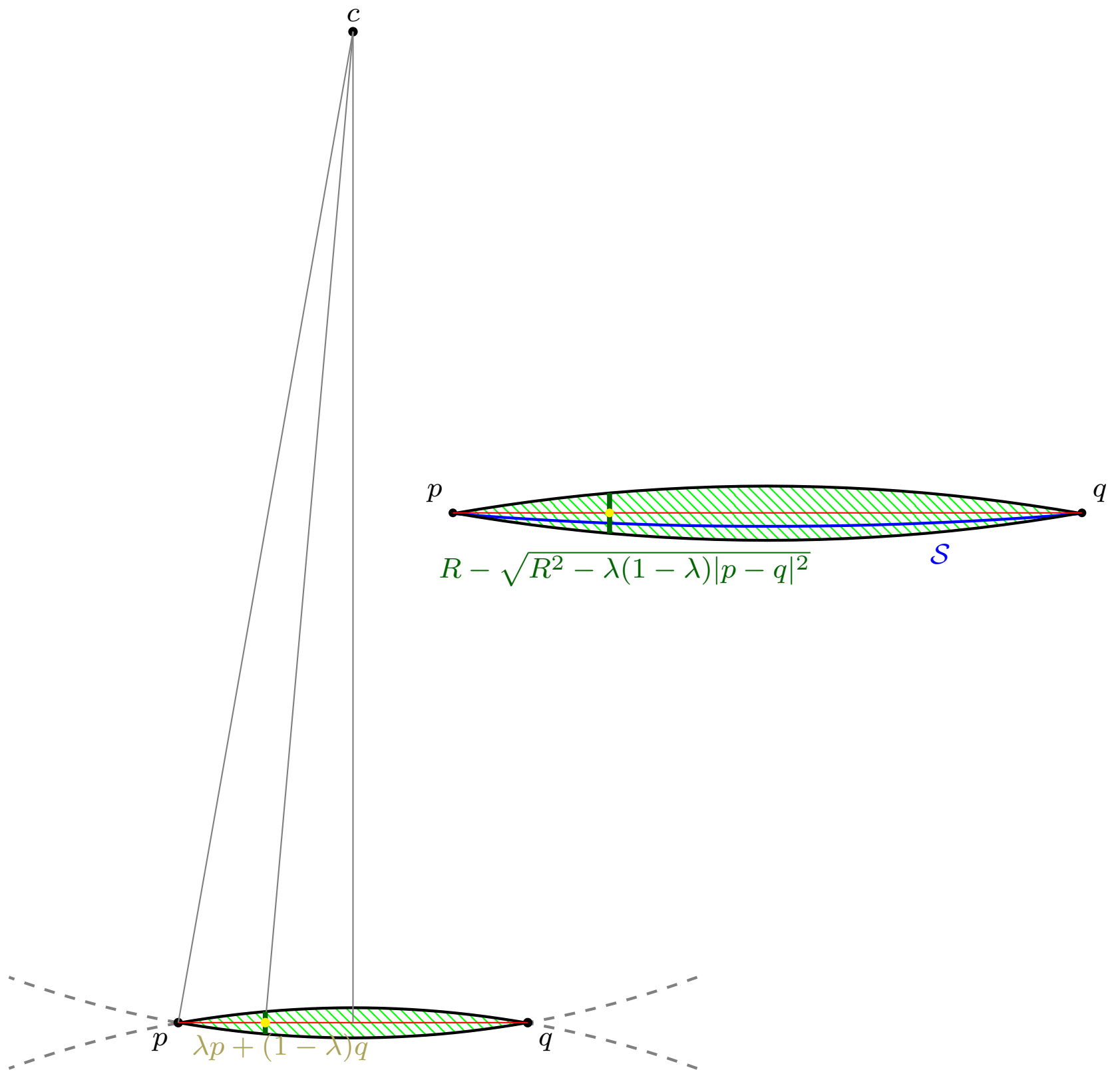
\begin{figure}[!ht]
    \centerline{
		\begin{minipage}{0.45\textwidth}
\begin{tikzpicture}
\coordinate [label={below left:$p$}] (p) at ({10*cos(260)}, {10*sin(260)+9.85});
\coordinate [label={below right:$q$}] (q) at ({10*cos(280)}, {10*sin(280)+9.85});
\fill [pattern=north west lines, pattern color=green, domain=260:280, variable=\x]
  plot ({10*cos(\x)}, {10*sin(\x)+9.85})
  -- cycle;
\fill [pattern=north west lines, pattern color=green, domain=80:100, variable=\x]
  plot ({10*cos(\x)}, {10*sin(\x)-9.85})
  -- cycle;
\draw [thick,domain=80:100] plot ({10*cos(\x)}, {10*sin(\x)-9.85});
\draw [thick,domain=260:280] plot ({10*cos(\x)}, {10*sin(\x)+9.85});
\draw [gray, thick, dashed, domain=70:80] plot ({10*cos(\x)}, {10*sin(\x)-9.85});
\draw [gray, thick, dashed, domain=260:250] plot ({10*cos(\x)}, {10*sin(\x)+9.85});
\draw [gray, thick, dashed ,domain=100:110] plot ({10*cos(\x)}, {10*sin(\x)-9.85});
\draw [gray, thick, dashed ,domain=290:280] plot ({10*cos(\x)}, {10*sin(\x)+9.85});
\draw[fill] (q) circle (0.04);
\draw[fill] (p) circle (0.04);
\coordinate [label={above:$c$}] (c) at (0,9.85);
\draw[fill] (c) circle (0.04);
\draw [red] (p)--(q);
\draw [gray] (c)--(p);
\draw [gray] (c)--({0}, {10*sin(280)+9.85});
\coordinate [label={below:{\color{yellow!60!black}$\phantom{33} \lambda p+ (1-\lambda) q$}}] (l) at ({10*cos(265)}, {10*sin(280)+9.85});
\draw [gray] (c)--(l);
\draw [line width=0.5mm, darkgreen] (l)--({10*cos(265)}, {10*sin(265)+9.85});
\draw [line width=0.5mm, darkgreen] (l)--({10*cos(95)}, {10*sin(95)-9.85});
\draw[fill,  yellow] (l) circle (0.04);
\end{tikzpicture}
\end{minipage}
\hspace{-1.8cm} 
\begin{minipage}{0.45\textwidth}
\begin{tikzpicture}[scale=1.8]
\coordinate [label={above left:$p$}] (p) at ({10*cos(260)}, {10*sin(260)+9.85});
\coordinate [label={above right:$q$}] (q) at ({10*cos(280)}, {10*sin(280)+9.85});
\fill [pattern=north west lines, pattern color=green, domain=260:280, variable=\x]
  plot ({10*cos(\x)}, {10*sin(\x)+9.85})
  -- cycle;
\fill [pattern=north west lines, pattern color=green, domain=80:100, variable=\x]
  plot ({10*cos(\x)}, {10*sin(\x)-9.85})
  -- cycle;
\draw [thick,domain=80:100] plot ({10*cos(\x)}, {10*sin(\x)-9.85});
\draw [thick,blue, domain=80:100] plot ({10*cos(\x)}, {10*sin(\x)-9.85+(\x-80)*(\x-100)/450});
\draw [thick,domain=260:280] plot ({10*cos(\x)}, {10*sin(\x)+9.85});
\draw[fill] (q) circle (0.02);
\draw[fill] (p) circle (0.02);
\draw [red] (p)--(q);
\coordinate  (l) at ({10*cos(265)}, {10*sin(280)+9.85});
\draw [line width=0.5mm, darkgreen] (l)--({10*cos(265)}, {10*sin(265)+9.85});
\draw [line width=0.5mm, darkgreen] (l)--({10*cos(95)}, {10*sin(95)-9.85});
\draw[fill,  yellow] (l) circle (0.02);
\coordinate [label={below:{\color{darkgreen}$\phantom{33} R - \sqrt{ R^2 - \lambda (1-\lambda) |p-q|^2} $}}] (l2) at ({10*cos(265)}, {10*sin(280)+9.70}); 
\coordinate [label={below:{\color{blue}$\phantom{33} \Su $}}] (l3) at ({10*cos(275)}, {10*sin(275)+9.85}); 
\end{tikzpicture}
\end{minipage}
    }
\caption{Left: The region in which the geodesic connecting $p$ and $q$ lies is indicated in green, the grey dashed circles have radius $R$. {Right: A zoom in on the spindle with the distance from the yellow point ($\lambda p+ (1-\lambda) q$) to the boundary of the spindle indicated in dark green. The geodesic connecting $p$ and $q$ (which is a part of $\Su$) within the spindle is indicated in blue. Because the union of the two dark green line segments (which are almost collinear in the figure, in higher dimensions it is a cone or in other words, the join of a sphere and a point), separates the spindle into two halves and the geodesic is continuous the dark green set intersect the geodesic. This implies that the distance from the grey point to the boundary upper bounds the distance to $\Su$.}    } 
\label{fig:lens}
\end{figure}

\begin{proof} 
Corollary \ref{corollary:LocalGeodesicConvexity} implies that the geodesic connecting $p$ and $q$ is contained in the intersection of all balls of radius less than the reach that contain $p$ and $q$ so that in particular the geodesic is contained in the spindle {(the intersection of all balls of radius $R$ in Euclidean space that contain $p$ and $q$)}, as depicted in Figure \ref{fig:lens}. This implies in turn that $|x - \pi _\Su (x) |$ less than or equal to the distance from $x$ to the {boundary of} the spindle. Let's restrict ourselves to the plane given by $p$, $q$, and $\pi _\Su (x)$ and consider a circle of radius $R$ with centre $c$ that yields the boundary of the spindle in this plane, see again Figure \ref{fig:lens}. By the cosine rule have that 
\[
| c-   \lambda p - (1 -\lambda) q|^2 = R^2 +  \lambda^2  |p - q|^2 - 2 R \lambda  |p - q| \cos \theta
\]
where $\cos \theta = \frac{| p-q| } {2R}$ (the value for $\theta$ follows by inspection of the triangle with vertices $p$, $c$, and $\frac{1}{2} (p+q)$). 
So that 
\begin{align} 
| c-   \lambda p - (1 -\lambda) q|^2 &= R^2 +  \lambda^2  |p - q|^2 - \lambda  |p - q|^2 
\nonumber 
\\ & = R^2 - (1-\lambda) \lambda  |p - q|^2  .
\nonumber
\end{align}  
The first inequality now follows by observing that the distance from $x$ to the spindle is given by 
$R - | c-   \lambda p + (1 -\lambda) q|$, see Figure \ref{fig:lens}. The second inequality follows from the standard observation that $\sqrt{1-t}= 1-\frac{t}{2} - \mathcal{O} (t^2)$, which in turn follows by Taylor's theorem. 
\end{proof}

We'll need the following corollary of \cite[Theorem 4.8(8)]{Federer}. Theorem 4.8(8) of \cite{Federer} says the closest point projection $\pi_\Su$ at a point $x$ is $\frac{1}{1- d(x, \Su) /R}$-Lipschitz. We therefore immediately have: 
\begin{corollary}\label{cor:LengthProjectedCurve} 
If $\gamma : [0 , \ell ]\to \mathbb{R}^d $ is a $C^1$ curve, then the length of $\pi_\Su(\gamma)$ is upper bounded as follows
\begin{align}
\length (\pi_\Su(\gamma)  )  \leq  \int_0 ^\ell  \frac{1}{1- d(\gamma(t) , \Su) /R}  |\dot{\gamma} (t)| d t .
\label{eq:UppBoundLenghtGen} 
\end{align} 
\end{corollary} 

The following simple but very useful bound will be instrumental in the proof below.  
\begin{lemma} \label{lem:Shortening} 
If $a$ and $b$ are two unit vectors we have that
\begin{align} 
|\lambda a+ (1-\lambda) b| ^2
&=  1- (1-\lambda)\lambda |a- b| ^2. 
\nonumber
\end{align}
For vectors of arbitrary length we have 
\begin{align} 
|\lambda a+ (1-\lambda) b| ^2 &= \lambda |a|^2 + (1-\lambda) |b|^2 - \lambda(1-\lambda) |a-b|^2 .
\nonumber 
\end{align} 
\end{lemma} 
\begin{proof} 
We have that for any two unit vectors $a,b$ that:
\begin{align} 
|\lambda a+ (1-\lambda) b| ^2 &=  \lambda^2 |a| ^2 + (1-\lambda)^2 |b|^2 + 2\lambda  (1-\lambda) a \cdot b
\label{eq:LSum1} 
\\ 
 &=  \lambda^2 + (1-\lambda)^2  + 2\lambda  (1-\lambda) a \cdot b
\tag{using that $|a|=|b|=1$}
\\ 
 &=  1- (1-\lambda)\lambda \left( 2  - 2 a \cdot b \right) 
\nonumber
\\ 
|a- b| ^2&= |a|^2+|b|^2 -2 a \cdot b
\label{eq:LSum2}
\\ 
& = 2- 2 a \cdot b
\tag{using that $|a|=|b|=1$}
\end{align}
and thus 
\begin{align} 
|\lambda a+ (1-\lambda) b| ^2
&=  1- (1-\lambda)\lambda |a- b| ^2. 
\nonumber
\end{align}
For $a,b$ of arbitrary length we combine \eqref{eq:LSum1} and \eqref{eq:LSum2} as follows
\begin{align} 
|\lambda a+ (1-\lambda) b| ^2 &= \lambda^2 |a| ^2 + (1-\lambda)^2 |b|^2 + \lambda  (1-\lambda) \left( |a|^2 + |b|^2 - |a-b| ^2 \right) 
\nonumber 
\\ 
&= \lambda |a|^2 + (1-\lambda) |b|^2 - \lambda(1-\lambda) |a-b|^2 .
\nonumber 
\end{align} 
\end{proof}


\UniqueG*


\paragraph*{Overview of the proof} $\phantom{4}$
The core idea of the proof is the following: Given two geodesics $\gamma_1$ and $\gamma_2$ connecting $a,b$ that are parametrized by arc length we have that $\lambda \gamma_1 + (1- \lambda) \gamma_2$ is shorter than $\gamma_1$ and $\gamma_2$ because $|\lambda \dot{\gamma}_1 + (1- \lambda) \dot{\gamma}_2 |<1$, if $\lambda \in (0,1)$. So\footnote{Here abuse notation a little bit because we do not distinguish the notation for a geodesic and its parametrization. } if $\lambda \gamma_1 + (1- \lambda) \gamma_2 \subset \Su$ we would find a contradiction, because there would be a curve connecting $a$ and $b$ that would be shorter than a geodesic. The main difficulty of the proof is that $\lambda \gamma_1 + (1- \lambda) \gamma_2$ is generally not in $\Su$, but we need to project on $\Su$ to find a curve that is. The technical difficulty lies with dealing with this projection. 

{The proof, perhaps surprisingly, also relies on a result from mathematical physics. 
Consider the Lagrangian of the {(real)} harmonic oscillator:
\[ \mathcal{L}_{\textrm{HO}} = |\dot{x}|^2- \frac{|x|^2}{R^2}
\]
Write $S_{{\textrm{HO}}}$ for the action, that is 
\begin{align} 
S_{{\textrm{HO}}}= \int_0 ^\tau  \mathcal{L}_{\textrm{HO}} dt . 
\label{eq:action}
\end{align} 
The Euler-Lagrange equation, 
\[
\frac{d}{dt} \frac{\partial \mathcal{L} } {\partial \dot{x} }= \frac{\partial \mathcal{L} }{\partial x},
\] 
in this case yields 
\begin{align} 
\ddot{x} (t) =- \frac{x(t)}{R^2}, 
\nonumber
\end{align}  
with the familiar solutions $c_1 \sin(t/R) + c_2 \cos (t/R) = c_4 \sin \left(\frac{t-t_0}{R} \right)$.
We have the following, see \cite[Section V]{Gray2007}, \cite{Moriconi2017} for this specific result\footnote{Note that our the Lagrangian of a high dimensional harmonic oscillator (with constant potential) is the sum of one-dimensional harmonic oscillators.} and \cite{gelfand2000calculus} for general theory: 
\begin{theorem}[Adjusted from \cite{Gray2007} and \cite{Moriconi2017}] \label{Thm:criticalOptimal} 
The solution to the Euler-Lagrange equation {of the harmonic oscillator}  is a global minimum of the action~\eqref{eq:action} provided that $\tau < \pi R $.
\end{theorem}
This means in particular that:
\begin{corollary} \label{Cor:criticalOptimal} 
Let $c_4$ be a constant. 
If $\tau  < \pi R $ and $x(t)$ is a function that satisfies $x(0)=0$ and $x(\tau)= c_4 \sin  \left(\frac{\tau}{R} \right)$, then we have that 
\begin{align}  
S_{{\textrm{HO}}} (x) &= \int_0 ^\tau   |\dot{x}|^2- \frac{|x|^2}{R^2}  dt  
\nonumber 
\\
&\geq \int_0 ^\tau    \left(\frac{ c_4^2 \cos^2 \left(\frac{t}{R} \right)  - c_4 ^2 \sin^2  \left(\frac{t}{R} \right) }{R^2}  \right)dt
\nonumber 
\\
&=\int_0 ^\tau    
\frac{ c_4^2}{R^2}    \cos \left(\frac{2t}{R} \right)    d t 
\nonumber
\\
& = \frac{ c_4^2}{2R} \sin \left(\frac{2 \tau}{R} \right).
\nonumber 
\end{align} 
\end{corollary} 
}

\begin{proof}[Proof of Theorem \ref{prop:Unique}] 
Because $d_S(a,b)< \pi R$ the geodesic distance between $a$ and $b$ is finite, which means that there exists a path of finite length connecting $a$ and $b$ in $\Su$.  Moreover, thanks to the second paragraph of part III, section
1: ``Die Existenz geod{\"a}tischer Bogen in metrischen R{\"a}ume'' in \cite{menger1930untersuchungen}, we have that: If there is at least one path between
$a$ and $b$ with finite length, then a minimizing geodesic, i.e. a path with
minimal length connecting $a$ to $b$, exists.  
Therefore the set {of minimizing geodesics} is not empty. 

Let  $\gamma_1,\gamma_2:[ 0,\ell ] \rightarrow S$ be two 
length minimizing paths
from $a$ to $b$,  parametrized by arc length,
with same length: $\length(\gamma_1) = \length(\gamma_2) = \ell := d_S(a,b)$.

Denote by $\epsilon$ the maximal distance between $\gamma_1 (t)$ and $\gamma_2(t)$, i.e. 
\begin{equation}\label{eq.DefEpsilon}
\epsilon :=  \max_{t\in [0, \ell]} | \gamma_1(t) -   \gamma_2(t)|, 
\end{equation}
and let $t_{\max} \in [0, \ell]$ be a parameter for this distance is attained, that is, 
\[
d \big( \gamma_1(t_{\max}),   \gamma_2(t_{\max}) \big) = \epsilon . 
\]
Without loss of generality, we assume that $t_{\max} \leq \ell /2$.


We now consider the two families of curves 
\[\gamma_\lambda ^{\mathbb{E} } (t) = \lambda \gamma_1 (t)+ (1-\lambda) \gamma_2 (t)
\] 
and
\[\gamma_\lambda ^{\Su} (t) =  \pi _{\Su} ( \gamma_\lambda ^{\mathbb{E} } (t)) . 
\] 
We now upper bound the length $\gamma_\lambda ^{\Su} (t)$. By Lemma \ref{lemDistProjMidpoint} we have that 
\begin{align} 
d( \gamma_\lambda ^{\mathbb{E} } (t), \Su ) \leq R - \sqrt{ R^2 - \lambda (1-\lambda) |\gamma_1(t) - \gamma_2(t)|^2}.
\nonumber
\end{align} 
Using Corollary \ref{cor:LengthProjectedCurve} 
this yields that 
\begin{align}
&\length (\gamma_\lambda ^{\Su} (t) )  
\nonumber
\\
&\leq  \int_0 ^\ell  \frac{1}{1- (1 - \sqrt{ 1 - \lambda (1-\lambda) |\gamma_1(t) - \gamma_2(t)|^2/R^2})   }  |\lambda \dot{\gamma}_1 (t)+ (1-\lambda) \dot{\gamma}_2 (t)| d t .
\label{eq:UppBoundLenght1} 
\end{align} 
We now focus on the second term in the product in the integral.

With Lemma \ref{lem:Shortening}  we find the following bound on \eqref{eq:UppBoundLenght1}: 
\begin{align}
\length (\gamma_\lambda ^{\Su} (t) )  \leq  \int_0 ^\ell  &\frac{\sqrt{ 1- (1-\lambda)\lambda (| \dot{\gamma}_1 (t)- \dot{\gamma}_2 (t)|^2 )}}{ \sqrt{ 1 - \lambda (1-\lambda) |\gamma_1(t) - \gamma_2(t)|^2/R^2}   } 
  d t .
\label{Int:Com1}
\end{align}
Assuming that $ (1-\lambda)\lambda$ is very small {(more precisely we'll show that the length decreases in first order near $\lambda= 0$ or $\lambda=1$)} yields that 
\begin{align} 
\int_0 ^\ell  &\frac{\sqrt{ 1- (1-\lambda)\lambda (| \dot{\gamma}_1 (t)- \dot{\gamma}_2 (t)|^2 )}}{ \sqrt{ 1 - \lambda (1-\lambda) |\gamma_1(t) - \gamma_2(t)|^2/R^2}   }   d t .
\nonumber
\\ =& \int_0 ^\ell \left (1+ \lambda (1-\lambda) \frac{|\gamma_1(t) - \gamma_2(t)|^2}{2 R^2} \right) \left(1 - \frac{1}{2} (1-\lambda)\lambda (| \dot{\gamma}_1 (t)- \dot{\gamma}_2 (t)|^2 )  \right)d t
\nonumber 
\\ 
&+ \mathcal{O} \left(\lambda^2 (1-\lambda)^2 \right )  
\tag{by Taylor}
\\ =& \int_0 ^\ell 1+ \lambda (1-\lambda) \frac{|\gamma_1(t) - \gamma_2(t)|^2}{2 R^2}   - \frac{1}{2} (1-\lambda)\lambda ( | \dot{\gamma}_1 (t)- \dot{\gamma}_2 (t)|^2 )   d t 
\nonumber 
\\ 
&+ \mathcal{O} \left(\lambda^2 (1-\lambda)^2 \right )  
\nonumber
\\ =& \int_0 ^\ell 1  - \frac{1}{2} (1-\lambda)\lambda  \left(| \dot{\gamma}_1 (t)- \dot{\gamma}_2 (t)|^2 -\frac{|\gamma_1(t) - \gamma_2(t)|^2}{R^2}  \right)   d t 
\nonumber 
\\ 
&+ \mathcal{O} \left(\lambda^2 (1-\lambda)^2 \right )  
\label{eq:Taylor} 
\end{align}

Now writing $\Delta (t) = \gamma_1(t) - \gamma_2(t) $ and thus $\dot{\Delta} (t) = \dot{\gamma}_1 (t)- \dot{\gamma}_2 (t)$, this integral (
ignoring the higher order terms) takes the form of a familiar Lagrangian integral or action 
\begin{align} 
\int_0 ^\ell \mathcal{L} \,  dt =
\int_0 ^\ell 1  - \frac{1}{2} (1-\lambda)\lambda  \left(| \dot{\Delta} (t)|^2 -\frac{|\Delta (t)|^2}{R^2}  \right)   d t ,  
\label{eq:LagrangianIntegral}
\end{align} 
which we recognize (up to global additive and multiplicative factors) as the action integral of the harmonic oscillator. 
To achieve better bounds we'll now split this into two integrals 
\begin{align} 
\int_0 ^\ell \mathcal{L} \,  dt = \int_0 ^{\ell/2}  \mathcal{L} \,  dt + \int_{\ell/2}^{\ell}  \mathcal{L} \,  dt.
\nonumber 
\end{align} 
We'll only concentrate on the first integral, the second one can be treated in a completely symmetrical manner. 

{We now invoke Theorem \ref{Thm:criticalOptimal} and Corollary \ref{Cor:criticalOptimal}, with $\tau = \ell/2$, together with \eqref{Int:Com1}, \eqref{eq:Taylor}, and \eqref{eq:LagrangianIntegral}, to find that 
\begin{align} 
\length (\gamma_\lambda ^{\Su} (t) )   
&\leq 
\int_0 ^\ell 1  - \frac{1}{2} (1-\lambda)\lambda  \left(| \dot{\gamma}_1 (t)- \dot{\gamma}_2 (t)|^2 -\frac{|\gamma_1(t) - \gamma_2(t)|^2}{R^2}  \right)   d t 
\nonumber 
\\ 
&\phantom{\leq} + \mathcal{O} \left(\lambda^2 (1-\lambda)^2 \right )  
\nonumber
\\
&= \left( \int_0 ^{\ell/2} + \int_{\ell/2}^\ell \right) \left( 1  - \frac{1}{2} (1-\lambda)\lambda  \left(| \dot{\gamma}_1 (t)- \dot{\gamma}_2 (t)|^2 -\frac{|\gamma_1(t) - \gamma_2(t)|^2}{R^2}  \right) \right)   d t 
\nonumber 
\\ 
&\phantom{\leq} + \mathcal{O} \left(\lambda^2 (1-\lambda)^2 \right )  
\nonumber
\\
&\leq 2 \left( \ell/2 - \frac{1}{2} (1-\lambda)\lambda  \frac{c_4^2}{2R} \sin \left(\frac{\ell}{R} \right)\right) + \mathcal{O} \left(\lambda^2 (1-\lambda)^2 \right )  , 
\tag{the $\leq$ follows because $- \frac{1}{2} (1-\lambda)\lambda$ is negative}
\end{align}   
where $c_4$ is some constant. 
%
%
This is strictly less than $\ell$, as long as $\ell <  \pi R$. This contradicts the assumption that the geodesics have length $\ell$ (which is the sum of the integral from $0$ to $\ell/2$ and the integral from $\ell/2$ to $\ell$), because now we found a curve that is shorter up to first order. }
\end{proof}

\subsection{Continuity of geodesics} 

\subsubsection{Continuity in the $C^0$ sense}

The uniqueness of the geodesics of length less than $\pi R$ implies continuity for geodesics under the same restriction. We will first prove a weak version of this statement, namely continuity without convergence of the derivatives. We will then proceed to prove that geodesics converge in the the $C^1$ sense, that is both the geodesics and its derivatives converge.
The proof of the continuity will rely heavily on the Arzel{\`a}-Ascoli theorem, see e.g. \cite{dunford1988linear}. To this end we have to recall some nomenclature: 
A sequence $\{\gamma_n\}_{n\in \mathbb{Z}_{\geq 0}}$ of continuous functions on an interval $I\subset \mathbb{R}$ is uniformly bounded if for all $n$ and all $t\in I$, we have that $|\gamma_n(x) | \leq C$, with $C$ some (universal) constant.  
The sequence $\{\gamma_n\}_{n\in \mathbb{Z}_{\geq 0}}$ is uniformly equicontinuous if, for every $\epsilon > 0$  there exists a $\delta>0$, such that for all $t,t'$ satisfying $|t-t'| \leq \delta$ and all $n$ we have that
\[|\gamma_n(t) - \gamma_n (t')| \leq \epsilon .\] 
The theorem can now be stated as
\begin{theorem}[Arzel{\`a}-Ascoli]
Let $\{\gamma_n\}_{n\in \mathbb{Z}_{\geq 0}}$ be a sequence of real-valued uniformly bounded, equicontinuous functions on a compact interval $I$. Then there exists a subsequence $\{\gamma_{n_k}\}_{k \in \mathbb{Z}_{\geq 0}}$ that converges uniformly. 
\end{theorem}

With these preliminaries out of the way we can state a weak form of the second contribution of the paper:
\begin{corollary}
\label{Cor:ContinuousG}  
Let $\Su$ be a set of positive reach $R$, then the geodesic connecting any two points $a,b\in \Su$ is continuous (in the $C^0$ sense) with respect to the endpoints provided that $d_\Su (a,b) <\pi R$. 
\end{corollary} 
In the next subsection this will be strengthened to continuity in the $C^1$ sense. 


 We provide two proofs, one here and another in the appendix. Although they share a fair number of common techniques they differ significantly on the way that they treat the length of the limit curve. Moreover, the proof here is shorter, but requires some familiarity with using the Arzel{\`a}-Ascoli theorem, while the proof in the appendix discusses this in greater detail. 
\begin{proof}
Let  $a_n$, $b_n$ be a pair of sequences of points in $\Su$, where $\lim_{n \to \infty} a_n =a$ and $\lim_{n \to \infty} b_n=b$, as well as a sequence of geodesics $\gamma_n$, connecting $a_n$ and $b_n$. Let $\gamma$ be the minimizing geodesic connecting $a$ and $b$.

We want to prove that $\lim_{n\rightarrow \infty} \| \gamma - \gamma_n\|_{\sup} = 0$,
 where $\| \cdot \|_{\sup}$ denotes the $C^0$ norm (i.e. sup norm).
 In other words, we have to prove that, for any  $\epsilon >0$, the cardinal of the set:
 \[
 \Gamma_\epsilon := \{\gamma_n \mid  \| \gamma - \gamma_n\|_{\sup} \geq \epsilon \}
 \]
 is finite.
 
 Assume to achieve a contradiction that the cardinal of $\Gamma_\epsilon$ is infinite.
Since the set of functions in $ \Gamma_\epsilon$ is bounded and equicontinuous (because they admit a uniform Lipschitz constant), 
we can apply Arzel{\`a}-Ascoli theorem and find a subsequence 
$\{\gamma_{n_k}, k\in \N\}\subset  \Gamma_\epsilon$ converging to a curve $\gamma'$ in $\Su$.
Since $|\textrm{length} (\gamma) - \textrm{length} (\gamma_{n_k}) | =|d_\Su (a,b) - d_\Su (a_{n_k},b_{n_k})| \leq d_\Su (a,a_{n_k}) + d_\Su (b,b_{n_k}). 
)$ one has $\lim_{k\rightarrow \infty} \textrm{length} (\gamma_{n_k}) = \textrm{length} (\gamma)$. 

By a classical result \cite[Claim 5.8, page 19]{Busemann2005},
the length of the limit cannot be larger than the limit inf of lengths which yields 
$\textrm{length} (\gamma') \leq \lim_{k \rightarrow \infty} \textrm{length} (\gamma_{n_k})  = \textrm{length} (\gamma)$.
Therefore, $\gamma'$ would be a geodesic between $a$ and $b$ with 
$\| \gamma - \gamma'\|_{\sup} \geq \epsilon$ which  contradicts the unicity of geodesics stated in Theorem \ref{prop:Unique}.
\end{proof}

{
Corollary \ref{Cor:ContinuousG} has itself another consequence which is worth pointing out. 
\begin{corollary}\label{Cor:CorollaryOfCorollary}
A geodesic ball of radius less than $\pi R$ is contractible. 
\end{corollary}  
\begin{proof} Write $B_{\Su} (c, r)$, for the geodesic ball in $\Su$ with centre $c$ and radius $r$, where $r < \pi R$.  Let $x \in B_{\Su} (c, r)$, and let $\gamma_{xc}(t)$ be the geodesic parametrized by $[0,1]$ with constant arclength with $\gamma_{xc}(1)=x$ and $\gamma_{xc}(0)=c$.
Thanks to Corollary \ref{Cor:ContinuousG} the geodesics are continuous under the $\sup$-norm, with respect to endpoint $x$. This means that the map $H: [0,1] \times B_{\Su} (c, r) : (\lambda,x) \mapsto \gamma_{xc}(\lambda)$ is a (continuous) homotopy. 
\end{proof} 
}

\subsubsection{Continuity in the $C^1$ sense}
{The continuity of the geodesics in the $C^1$ sense is a consequence of the fact that geodesics in a set of positive reach have positive reach, due to the metric characterization of sets of positive reach \cite[Theorem 1]{TangentVarJournal}. This means \cite{FullVerionc11, lytchak2004geometry, lytchak2005almost} that the geodesics are $C^{1,1}$ and moreover Lipschitz constant of the first derivative is upper bounded by $1/R$ (with respect to the arclength), with $R$ the reach. In particular this means if two geodesics of the same reach have a large angle between the tangent vectors (at some point along the curves)  then the two geodesics cannot be close (in the $C^0$ or the Hausdorff) sense everywhere in neighbourhood of the point, because the geodesics cannot bent inward fast enough, to stay close to each other.    
}

\paragraph{General vector valued functions}
The link between the distance in the $C^0$ and $C^1$ senses for a $C^{1,1}$ curve can be formalized based on the distance function. Indeed we have the following: 
\begin{lemma}\label{lemma.GeneralC11VectorValuedFunction}
Let $F: [0,c]  \rightarrow \R^d$, be a $C^1$ function whose derivative $\dot F$ is $\varepsilon_2$-Lipschitz.
We write $\varepsilon_0$ and $\varepsilon_1$ for the supremum norms on the function and its derivative
\[
\varepsilon_0 := \sup_{t\in [0,c] } | F(t) |, \quad \varepsilon_1 :=  \sup_{t\in [0,c] } | \dot F(t) |. 
\]
With this notation one has,
\begin{equation}\label{eq.C1BoundedByC0GeneralFunctions}
\varepsilon_1 \leq \max \left( \sqrt{4 \varepsilon_0 \varepsilon_2},   4  \varepsilon_0 / c \right).
\end{equation}
\end{lemma}

\begin{proof}
Since $t \mapsto | \dot F(t)|$ is continuous, there exists $t_{\max} \in [0,c]$ such that
\[
\varepsilon_1 = | \dot F(t_{\max}) |. 
\]
Define $[a,b] \subset [0,c]$ based on two cases: 
\begin{itemize} 
\item If $\varepsilon_2 c \leq \varepsilon_1$ we pick $[a,b] = [0,c]$. 
\item If $\varepsilon_2 c \geq \varepsilon_1$ we choose any $a,b$, such that $t_{\max} \in [a,b]$ and $b-a = \frac{\varepsilon_1}{\varepsilon_2}$.
\end{itemize} 
Observes that, in both cases we have that 
$t_{\max} \in [a,b]$ and
\[
b-a \leq \frac{\varepsilon_1}{\varepsilon_2}.
\]
We can assume that $\varepsilon_1 \neq 0$ as then the Lemma trivially holds. We now define
\[
T := \frac{\dot F(t_{\max})}{| \dot F(t_{\max}) | }. 
\]
One has,
\begin{align}
2 \varepsilon_0 &\geq | F(a) | + |F(b)| 
\nonumber
\\
&\geq | F(b) - F(a) | 
\nonumber
\\
& = \left|  \int_a^b  \dot F(t) dt \right| 
\nonumber
\\
& \geq \left \langle T,  \int_a^b  \dot F(t) dt \right \rangle 
\nonumber
\\
&=  \int_a^b   \langle T,  \dot F(t) \rangle dt  
\nonumber
\\
&\geq \int_a^b | \dot F(t_{\max}) | - \varepsilon_2 | t - t_{\max} | dt  
\tag{since $\left| \frac{d}{dt} \langle T,  \dot F(t)  \rangle dt \right| \leq \varepsilon_2$} 
\\
& 
\geq (b-a)  \varepsilon_1 -  \frac{1}{2} \varepsilon_2 (b-a)^2 
\nonumber
\\
& \geq  \frac{1}{2} (b-a)  \varepsilon_1. \tag{since $b-a \leq \frac{\varepsilon_1}{\varepsilon_2}$} 
\end{align}
We now come back to the two cases used in the definition of $[a,b]$: In the first case, where $\varepsilon_2 c \leq \varepsilon_1$, then  $b-a = c$ and one gets $\varepsilon_1 \leq 4 \varepsilon_0 / c$. 
In the second case, where $\varepsilon_2 c \geq \varepsilon_1$, then $b-a = \frac{\varepsilon_1}{\varepsilon_2}$ and we get $\frac{1}{2} \frac{\varepsilon_1^2}{\varepsilon_2} = 2 \varepsilon_0$
which gives $\varepsilon_1 \leq  \sqrt {4 \varepsilon_0 \varepsilon_2}$.
\end{proof}

\paragraph{Curves are continuous in the $C^1$ sense}

We can now draw the conclusions from Lemma \ref{lemma.GeneralC11VectorValuedFunction}, which we formulate in a number of different ways, with an eye on future applications. 

\paragraph{Curves with same length}
In the first case we compare two curves (or geodesics) of the same length. 
\begin{corollary}
Let $\gamma_i: [0,\ell] \rightarrow \R^d$, for $i=1,2$ be curves of same length $\ell >0$ parameterized by arc length and with reach at least $R>0$.
Then, denoting
\[
\varepsilon_0 := \sup_{s\in [0,\ell] } |  \gamma_2(s)  -  \gamma_1(s)  |, \quad \varepsilon_1 := \sup_{s\in [0,\ell] } | \dot \gamma_2(s) - \dot \gamma_1(s) |
\]
one has:
\begin{equation}\label{eq.C1BoundedByC0}
\varepsilon_1 \leq \max \left( \sqrt{8 \varepsilon_0 / R },   4  \varepsilon_0 / \ell \right)
\end{equation}
\end{corollary}
\begin{proof}
We use Lemma~\ref{lemma.GeneralC11VectorValuedFunction} with $F:= \gamma_2 - \gamma_1$. For this we first need to establish the Lipschitz constant on $F$. 
As mentioned in the introduction to this section, the works \cite{FullVerionc11, lytchak2004geometry, lytchak2005almost} imply that the Lipschitz constant of the first derivative by is bounded by $1/R$ (with respect to the arclength), with $R$ the reach. This means that $\varepsilon_2 = 2/R$ is a Lipschitz constant for the derivative of the difference of two such curves.  
%
%
%
%
 %
 %
The result now follows. 
\end{proof}

\paragraph{Curves with uniform parameterization on $[0,1]$} In the second case we compare two curves (or geodesics) uniformly parametrized on $[0,1]$. 

\begin{corollary}
Let $\gamma_i: [0, 1] \rightarrow \R^d$, for $i=1,2$ be curves with uniform parameterization (i.e. $|\dot \gamma_i (t) | =  \length(\gamma_i)$ for any $t \in [0,1]$)
and with reach at least $R>0$.
Then, denoting
\begin{align*}
\varepsilon_0 &:= \sup_{t\in [0, 1] } |  \gamma_2(t)  -  \gamma_1(t)  |, \\
\quad \varepsilon_1 &:= \sup_{t\in [0, 1] } | \dot  \gamma_2(t)  - \dot  \gamma_1(t)  |, \\
\ell_i &:=  \length(\gamma_i),
\end{align*}

one has:
\begin{equation}
\varepsilon_1 \leq \max \left( \sqrt{4 (\ell_1^2 + \ell_2^2) \varepsilon_0 / R },   4  \varepsilon_0 \right)
\end{equation}
\end{corollary}

\begin{proof}
We again apply Lemma \ref{lemma.GeneralC11VectorValuedFunction} with $F:= \gamma_2 - \gamma_1 $.
Note that in this case, using the same argument as in the previous corollary, one has $\varepsilon_2 = (\ell_1^2 + \ell_2^2)/R$ as a Lipschitz constant on $| \dot F|$.
\end{proof}

We summarize, first in a general then a more specific corollary: 
\begin{corollary}
Let $\gamma [0, 1] \rightarrow \R^d$ and $\gamma_i: [0, 1] \rightarrow \R^d$, for $i\in \N$ be curves with uniform parameterization 
and with reach at least $R>0$ such that $(\gamma_i)_{i\in \N}$ converge to $\gamma$ in $C^0$ norm and the lengths of $\gamma_i$
are uniformly bounded, then $(\gamma_i)_{i\in \N}$ converge to $\gamma$ in $C^1$ norm.
\end{corollary}
The second more specific, is also the second main result of this paper:
\ContinuousGCOne*

\section{Sets of reach \texorpdfstring{$R$}{R} are \texorpdfstring{$\operatorname{CAT}\left( 1/R^2\right)$}{CAT 1/R2}}

\subsection{Notation} \label{sec:notation} 
In the second part of the paper geodesics will be arc length parametrized, so that a geodesics $\gamma_{a,b}$ is parametrized as
 \begin{equation*}
 \gamma_{a,b}: \left[0, \,  d_{\Set}(a,b) \right] \rightarrow \Set,
 \end{equation*}
with   $\gamma_{a,b}(0) = a$ and $\gamma_{a,b}( d_{\Set}(a,b))=b$.
We stress that as we have seen in Corollary~\ref{cor:TwiceDiffAlmostAnywhere} second derivative of $\gamma_{a,b}$ at $s$, denoted $\ddot{\gamma}_{a,b}(s)$, is exists almost everywhere.

Recall that arc length parametrization implies $|\dot{\gamma}_{a,b}(s)|=1$ and $\langle \dot{\gamma}_{a,b}(s), \ddot{\gamma}_{a,b}(s) \rangle = 0$.
Moreover, by Corollary \ref{lemma.geodesicExistPartA}, geodesics in $\Set$ have reach at least $R$, and in this particular setting Theorem 4.8(7) of \cite{Federer} reduces to:
\begin{lemma}\label{lemma.thisLemma}
If $a,b \in \Set$ satisfy $d_{\Set}(a,b)< \pi R$, then
\begin{equation}\label{eq.angleTgtGeodesicEuclideanSegment}
\sin \angle \dot{\gamma}_{a,b}(0) , b-a \leq \frac{|b-a|}{2R}. 
\end{equation}
\end{lemma}



{ 
Although we recalled the concept in the introduction, we give a more formal definition of comparison triangle and $\operatorname{CAT} (k)$ spaces here. We follow \cite{bridson2013metric} closely. 

A geodesic triangle $\Delta$ in a metric space $X$ consists of three vertices $\{ a,b,c\}$ and (a choice of) three (minimizing) geodesics joining the points pairwise, which are referred to as edges. A triangle $\bar{\Delta}$ with vertices $\{ \bar{a} , \bar{b} ,\bar{c} \}$ on a space form $\mathbb{H}(k)$ is called a comparison triangle for $\{ a,b,c\}$ if $d_X(a,b) = d_{\mathbb{H}(k)} ( \bar{a} , \bar{b} )$, $d_X(b,c) = d_{\mathbb{H}(k)} ( \bar{b} , \bar{c} )$, and $d_X(a,c) = d_{\mathbb{H}(k)} ( \bar{a} , \bar{c} )$. Such triangles exist if the perimeter $d_X(a,b)+ d_X(b,c)+d_X(a,c)$ is less than the diameter of $\mathbb{H}(k)$. Given a point $x$ on the geodesic in the geodesic triangle connecting two vertices, say $a$ and $b$ so that the distance to the vertices are $d_X (a,x)$, $d_X (b,x)$ respectively, then its comparison point $\bar{x}$ is the point on the corresponding geodesic which has the same distance two corresponding vertices, i.e.  $d_X (a,x)= d_{\mathbb{H}(k)}  (\bar{a},\bar{x})$,  $d_X (b,x)= d_{\mathbb{H}(k)}  (\bar{b},\bar{x})$. 

\begin{remark} 
For the general definition, we denote space forms, i.e. simply connected spaces of constant curvature (Euclidean space, hyperbolic spaces, or spheres), with $\mathbb{H}(k)$. In the specific setting of positive curvature, as we'll find ourselves in below, we prefer to specifically write $\Sphere^d$ for spheres. We stress that (as usual) the diameter of non-positively curved spaces is infinite. 
\end{remark} 

\begin{definition}[{Chapter II.1, Definition 1.1 \cite{bridson2013metric}}]
Let $X$ be a metric space and let $k\in \mathbb{R}$. Let $\Delta$ be a geodesic triangle whose perimeter is smaller than twice the diameter of $\mathbb{H}(k)$. Let $\bar{\Delta} \subset \mathbb{H}(k)$ be a comparison triangle for $\Delta$. Then, $\Delta$ is said to satisfy the $\CAT(k)$ inequality if for all $x,y \in \Delta$ and their comparison points $\bar{x},\bar{y} \in \bar{\Delta}$, we have that $d_X (x,y) \leq d_{\mathbb{H}(k)} (\bar{x} ,\bar{y})$. 
If  $k \leq 0$ then $X$ is a $\CAT(k)$ space or $X$ is $\CAT(k)$ if $X$ is a geodesic space all of whose geodesic triangles satisfy the $\CAT(k)$ inequality. 
If  $k > 0$, then $X$ is called a $\CAT(k)$ if all pairs of points that are less than the diameter of $\mathbb{H}(k)$ apart are joined by a geodesic and all geodesic triangles of perimeter less than that of twice the  diameter of $\mathbb{H}(k)$ satisfy the $\CAT(k)$ inequality.   
\end{definition}

We now recall the characterization of $\CAT(k)$ spaces that plays a central role in the second part of this paper. 
\begin{proposition}[{Adjusted from Chapter II.1, Proposition 1.7, Claims (1) and (4) \cite{bridson2013metric}}] 
Fix $k \in \mathbb{R}$. Let $X$ be a metric space such that all pairs of points that are less than the diameter of $\mathbb{H}(k)$ apart are joined by a geodesic. Then the following are equivalent: 
\begin{itemize} 
\item[(1)] $X$ is $\CAT(k)$.
\item[(4)] The angle between the sides of any geodesic triangle in $X$ (where we assume that the perimeter of each geodesic triangle is is smaller than twice the diameter of $\mathbb{H}(k)$ if $k>0$) with distinct vertices is no greater than the angle between the corresponding sides of its comparison triangle in $\mathbb{H}(k)$.  
\end{itemize} 
\end{proposition} 

\begin{remark} 
In the general definition the angle is specified to be the Alexandrov angle, see \cite[Chapter I.1 Definition 1.12]{bridson2013metric}, but in our setting it reduces to the usual Euclidean angle. 
\end{remark} 
}

\subsection{Overview of the proof}

The proof of the fact that sets of reach at least $R$,
 equipped with the inner geodesic distance,
 are $\operatorname{CAT} (1/R^2)$ proceeds in 4 main steps. 

\paragraph{Step 1} Whenever it exists $\ddot{\gamma}_{a,b}(s) \in \Nor_{\Set}( \gamma_{a,b}(s))$. 

\paragraph{Step 2} Given 
two geodesics $\gamma_1$, $\gamma_2$ of length $< \pi R$ emanating from $a$ with some angle between them, we prove that their Euclidean distance $s\mapsto | \gamma_2(s) - \gamma_1(s)|$ at a same length $s$ is not less  than the corresponding Euclidean distance between geodesics on a sphere of radius $R$ starting with the same angle (within the context of Riemannian geometry this is called a Rauch hinge). This proof proceeds in two stages.

Based on Step 1 we can prove the following differential inequality is satisfied for any pair of geodesics $\gamma_1, \gamma_2$:
\begin{equation}
 \langle \ddot \gamma_2(s_2) - \ddot \gamma_1(s_1), \gamma_2(s_2) - \gamma_1(s_1)  \rangle 
 \geq  - \frac{(\gamma_2(s_2) - \gamma_1(s_1))^2}{R^2}
\tag{\ref{eq:DiffInequalityGamma1Gamma2}}, 
\end{equation}
{whenever  $\gamma_1$ and $\gamma_2$ are twice differentiable at respective parameters $s_1$ and $s_2$.}
We apply \eqref{eq:DiffInequalityGamma1Gamma2} with $\gamma_i= \gamma_{a,b_i}$ for $i=1,2$
and $s_1=s_2= s$, where $s \in [0,\ell]$ and $\ell$ less than the minimal length of $\gamma_1$ and $\gamma_2$.
By defining: 
\[ 
r (s) := |\gamma_2(s) - \gamma_1(s) |
\] 
the inequality \eqref{eq:DiffInequalityGamma1Gamma2} reduces then to 
\begin{equation}
\ddot r \;+\;\frac{1}{R^2}r \;\ge\;0, \quad\text{with} \quad r(0) = 0 \,.
\tag{\ref{eq.TheDifferentialInequation}} 
\end{equation}
This differential inequality yields the following results for $r$: 
\begin{restatable}{lemma}{DifferentialInequalityONE}
\label{lemma.DifferentialInequality_1}
If $r: \R^+ \rightarrow \R$ is a $C^{1,1}$ function that satisfies \eqref{eq.TheDifferentialInequation}, then
 \begin{equation}\label{eq.DifferentialInequality_1}
 \forall t \in [0, \pi R], \: r(t) \geq R\dot r(0)\sin\frac{t}{R}.
 \end{equation}
\end{restatable}
Equation \eqref{eq.DifferentialInequality_1}  indeed means that the Euclidean distance between the points is bounded below by the corresponding distance for a sphere. 

\begin{remark} We stress that, while this relation is tight  for the Euclidean distance,  this is not an intrinsic quantity and this is not sufficient to derive the $\operatorname{CAT} (k)$ property in and by itself, even for isosceles triangles. The lower bound on the Euclidean distance does not yield a sufficiently strong lower bound on the geodesic distance between the points. 
\end{remark} 


\paragraph{Step 3}  
We construct a lemma that is reminiscent of the Gauss' lemma \cite[Lemma 3.5]{docarmo1992}
in Riemannian geometry.\footnote{The Gauss' lemma (which we refer to) says that any sufficiently small sphere centered at a point in a Riemannian manifold is perpendicular to every geodesic through the point. }
Roughly speaking we prove the boundary of a geodesic ball of radius $< \pi R$, is `orthogonal' to the radial geodesic.  
In a precise sense we have the following lemma:
\begin{restatable}{lemma}{GeodesicSphereOrthogonalToRadialGeodesics}
\label{lemma.GeodesicSphereOrthogonalToRadialGeodesics}
Given $0< \ell< \pi R$ , $\varepsilon>0$, and $a,b \in \Set$ with $d_{\Set}(a,b)=\ell$,
there exists $\alpha>0$  such that  for any $b' \in \Set$ that satisfies
$d_{\Set}(a,b')=\ell$  
and $0<|b' -b| < \alpha$, one has
\begin{equation}\label{eq.GeodesicSphereOrthogonalToRadialGeodesics}
\frac{\pi}{2} - \varepsilon \leq  \angle \,  b'-b,  \dot \gamma_{a,b} (\ell)  \leq \frac{\pi}{2} + \varepsilon. 
\end{equation}
\end{restatable}

 
\paragraph{Step 4}
We define what we call a unit inverse exponential map $\ulog_a$, where $a \in \Set$. This map takes a point $x \neq a $ in the open geodesic ball of radius $\pi R$ centered at $a$ 
and gives $\dot \gamma_{a,x}(0)$, the unit tangent vector to the geodesic from $a$ to $x$ at $a$. 
By the result of Step 2 we prove that this map is (locally) Lipschitz.

Now we consider three points $a,b,c \in \Set$, such that the perimeter of the geodesic triangle satisfies $d_\Set (a,b) +d_\Set (b,c)+d_\Set (a,c) < 2 \pi R$ and we want to upper bound the angle at the vertex $a$, which is the angle between $ \ulog_a(b)$ and $\ulog_a(c)$. 

We consider the composition of $\operatorname{ulog}_a$ with the geodesic $\gamma_{b,c}$ from $b$ to $c$, which is a Lipschitz curve on the unit ball and we remark that the length of this curve provides an upper bound  on the angle $\angle \ulog_a(b),\ulog_a(c)$. Let's call $\rho(s)$ the geodesic distance between $a$ and $\gamma_{b,c}(s)$, where  $\rho(s) <\pi R$ thanks to the triangle inequality.

Using Steps $2$ and $3$, we prove that this length is upper bounded as follows 
\begin{equation}
\angle \ulog_a(b),\ulog_a(c)
\leq 
\length( \ulog_a \circ \gamma_{b,c} ) \leq \int_0^{d_{\Set}(b,c)} \frac{\sqrt{1 - \dot \rho(s)^2} }{R \sin \frac{\rho(s)}{R} } ds.
\tag{\ref{eq.IntegralBoundInnerAngle}} 
\end{equation}
We consider the same relation on the standard 
$2$-sphere $R \Sphere^{2}$ of radius $R$, for a geodesic triangle 
$\bar a, \bar b,\bar c \in R \Sphere^{2}$ with the same edge lengths as  $abc$. We get: 
\begin{equation}
\angle \ulog_{\bar{a}}(\bar{b}),\ulog_{\bar{a}}(\bar c) =\length( \ulog_{\bar{a}} \circ \bar \gamma_{\bar{b},\bar{c}} ) = \int_0^{d_{R\Sphere^2}(\bar{b},\bar{c})} \frac{\sqrt{1 - \dot { \bar \rho}(s)^2} }{R \sin \frac{\bar \rho(s)}{R} } ds.
\nonumber
\end{equation}
Moreover in the case of the geodesic $\bar{\gamma}_{\bar{b},\bar{c}}$ on the sphere we prove that the corresponding $\bar{\rho}$ maximizes the integral under the constraints that:
\[
\bar \rho(0) = d_{R \Sphere^2 }(\bar a, \bar b) \quad \text{and} \quad \bar \rho(d_{R \Sphere^2}(\bar b, \bar c) ) = d_{R \Sphere^2}(\bar a,\bar c).
\]

\paragraph{Wrap-up} We have shown that the angle $\angle \ulog_a(b),\ulog_a(c)$ at $a$ in $\Set$ is upper bounded by the length of the curve $\ulog_a \circ \gamma_{b,c}$, which is itself upper bounded by \eqref{eq.IntegralBoundInnerAngle}, whose right hand side is upper bounded by the corresponding integral for $R\Sphere^2$, which is angle at $\bar{a}$ on the sphere. This in fact gives us the characterization of a $\operatorname{CAT}(1/R^2)$ set.    

\subsection{Geometry of geodesics in a set of positive reach}
It is well known that, thanks to the variational principle, the second order derivative of a geodesic is normal to a smooth submanifold in $\mathbb{R}^d$, see e.g. \cite{SpivakI}. This statement has the following generalization in the context of sets of positive reach. 
\begin{restatable}{lemma}{SecondDerivaticeIInNor}
\label{lemma.SecondDerivaticeIInNor}
If $a,b \in \Set$ satisfy $d_{\Set}(a,b)< \pi R$ and $0 < s  < d_{\Set}(a,b) $,
and the geodesic $\gamma_{a,b}$ is twice differentiable at $s$ then,
\begin{equation}\label{eq.SecondDerivativeInNor}
\ddot{\gamma}_{a,b}(s) \in \Nor_{\Set}( \gamma_{a,b}(s) ) . 
\end{equation}
Here, $\Nor_{\Set}(x)$ denotes the cone normal to $\Set$ at $x$. 
More generally, the Clarke derivative \cite{Clarke1990} of $s \mapsto \dot{\gamma}_{a,b}(s)$ at $s$ is a subset of $\Nor_{\Set}( \gamma_{a,b}(s))$, that is, 
\begin{align}\label{eq.ClarkeSecondDerivativeInNor}
\forall  s&  \in ( 0, d_{\Set}(a,b)), &  \partial \dot{\gamma}_{a,b}(s) & \subset \Nor_{\Set}( \gamma_{a,b}(s) ).
\end{align}
\end{restatable}
The proof is given in Appendix \ref{section.ProofLemmaSecondDerivaticeIInNor}.

\subsection{Geometry of a geodesic isosceles wedge}
\subsubsection{A second order differential inequality on pairs of geodesics}
Recall that, since geodesics are $C^{1,1}$ they are twice differentiable almost everywhere.
\begin{lemma}\label{lemma:DiffInequalityGamma1Gamma2}
For $i=1,2$, let $a_i,b_i \in \Set$ such that $d_{\Set}(a_i,b_i)< \pi R$, denote $\gamma_i:= \gamma_{a_i,b_i}$.
Let $s_i \in ( 0, d_{\Set}(a_i,b_i) )$ be such that $\gamma_i$ is twice differentiable at $s_i$.
One has that, 
\begin{equation}\label{eq:DiffInequalityGamma1Gamma2}
 \langle \ddot \gamma_2(s_2) - \ddot \gamma_1(s_1), \gamma_2(s_2) - \gamma_1(s_1)  \rangle 
 \geq  - \frac{|\gamma_2(s_2) - \gamma_1(s_1)|^2}{R^2}. 
\end{equation}
\end{lemma}
\begin{proof} 
 Corollary \ref{cor:TwiceDiffAlmostAnywhere} says that  $|\ddot \gamma_i | \leq 1/R$,  which means that \eqref{eq:DiffInequalityGamma1Gamma2} 
is trivially satisfied if $| \gamma_2(s_2) - \gamma_1(s_1) | \geq 2R$.

We assume now that $| \gamma_2(s_2) - \gamma_1(s_1) | < 2R$.
In this case, thanks to Corollary~\ref{lemma.geodesicExistPartA}, there exists a geodesic $\delta$ connecting $\gamma_1(s_1)$ 
and $\gamma_2(s_2)$,
\[\delta := \gamma_{\gamma_1(s_1), \gamma_2(s_2)}.
\]

One has,
\begin{align}
\dot \delta(0) &\in \Tan_{\Set} (\gamma_1(s_1))
\nonumber
\\
\ddot \gamma_1(s_1) &\in \Nor_{\Set} (\gamma_1(s_1)). \tag{by Lemma \ref{lemma.SecondDerivaticeIInNor} and \eqref{eq.SecondDerivativeInNor}}  
\end{align}
By definition of $\Nor_{\Set}$ one has then $\angle \ddot \gamma_1(s_1), \dot \delta(0) \geq \pi/2$, 
so that, 
by the triangle inequality for angles or equivalently the triangle inequality for points on the sphere,
\[
\angle \ddot \gamma_1(s_1), \gamma_2(s_2) - \gamma_1(s_1) \geq \pi/2 - \angle \dot \delta(0), \gamma_2(s_2) - \gamma_1(s_1).
\]
This, with \eqref{eq.angleTgtGeodesicEuclideanSegment}, gives us
 \[
  \cos \angle \ddot \gamma_1(s_1), \gamma_2(s_2) - \gamma_1(s_1) \leq  \frac{ | \gamma_2(s_2) - \gamma_1(s_1) |}{2R}, 
\]
so that
\begin{align*}
 \langle \ddot \gamma_1(s_1), \gamma_2(s_2) - \gamma_1(s_1)  \rangle 
 &\leq |\ddot \gamma_1(s_1)|  | \gamma_2(s_2) - \gamma_1(s_1) | \cos \angle \ddot \gamma_1(s_1), \gamma_2(s_2) - \gamma_1(s_1) \\
 & \leq  \frac{(\gamma_2(s_2) - \gamma_1(s_1))^2}{2R^2}.
\tag{since  $ |\ddot \gamma_1(s_1)| \leq 1/R$} 
\end{align*}
Swapping $\gamma_1$ and  $\gamma_2$ and adding the corresponding inequalities yields \eqref{eq:DiffInequalityGamma1Gamma2}.
\end{proof}

\subsubsection{Differential inequality for the divergence of a geodesic wedge with respect to the Euclidean distance}
We consider now $a,b_1,b_2 \in \Set$ such that $d_{\Set}(a,b_i)< \pi R$, for $i=1,2$.
We apply \eqref{eq:DiffInequalityGamma1Gamma2} with $\gamma_i= \gamma_{a,b_i}$ for $i=1,2$
and $s_1=s_2= s$, where $s \in [0,\ell]$ and $\ell$ the minimal length of $\gamma_1$ and $\gamma_2$.

We consider  $F: [0,\ell] \rightarrow \R^d$ defined by $F:=\gamma_2 -\gamma_1$ so that \eqref{eq:DiffInequalityGamma1Gamma2}  gives:
\begin{equation}\label{eq:DiffInequation}
\langle \ddot F(s), F(s) \rangle \geq -\frac{F(s)^2}{R^2} \quad\text{and} \quad F(0) = 0.
\end{equation}
\subsubsection{Geodesics on a sphere with radius $R$}\label{section:onTheSphere}
For $\psi \in [0, \pi]$ and $s\in [0, \pi R]$ we consider ``meridian'' geodesics $\bar \gamma_1$ and $\bar \gamma_2$ parametrized by arc length 
on the sphere of radius $R$ in $\R^3$, starting at the ``north pole''
 $(0,0,R)$, and defined by:
\begin{equation*}
\bar \gamma_1(s) := 
\begin{pmatrix}
R \cos \frac{\psi}{2} \sin \frac{s}{R} \\ 
- R \sin \frac{\psi}{2} \sin \frac{s}{R}   \\ 
R \cos \frac{s}{R} 
\end{pmatrix}, \:
\bar \gamma_2(s) := 
\begin{pmatrix}
R \cos \frac{\psi}{2} \sin \frac{s}{R}  \\ 
R \sin \frac{\psi}{2} \sin \frac{s}{R}   \\ 
R \cos \frac{s}{R} 
\end{pmatrix} \\
\end{equation*}

Observes that $\angle \dot {\overline \gamma}_1 (0),  \dot {\overline \gamma}_2 (0) = \psi$

Setting $\bar F (s) := \bar \gamma_2(s) - \bar \gamma_1(s)$ we get:
\begin{equation*}
\bar F (s) = 
\begin{pmatrix}
0\\ 
2 R \sin \frac{\psi}{2} \sin \frac{s}{R}   \\ 
0
\end{pmatrix}, \:
\dot {\bar F} (s) = 
\begin{pmatrix}
0\\ 
2 \sin \frac{\psi}{2} \cos \frac{s}{R}   \\ 
0
\end{pmatrix}, \:
\ddot {\bar F} (s) = 
\begin{pmatrix}
0\\ 
\frac{-2}{R}\sin \frac{\psi}{2} \sin \frac{s}{R}   \\ 
0
\end{pmatrix}
\end{equation*}
Observe that the function $\bar F$ satisfies the  equality in  inequality \eqref{eq:DiffInequation}
which is therefore tight.

\subsubsection{Differential inequality}
In this section, we infer a lower bound on $|F(s)|$ from \eqref{eq:DiffInequation}.

Denote $r(s):=|F(s)|$ 
and assume that $F\neq 0$. 
 Using that 
\begin{align*} 
\frac{d^2}{ ds^2} r^2 &= 2 \dot{r}^2 + 2 r \ddot{r}= \frac{d^2}{ ds^2} \langle F , F \rangle = 2 \langle \dot{F} ,\dot{F}\rangle + 2 \langle \ddot{F} ,F \rangle  
\\
\dot{r} &= \frac{d}{ ds} \sqrt{ \langle F , F \rangle } = \frac{\langle \dot{F} , F \rangle}{\sqrt{ \langle F , F \rangle }},
\end{align*} 
 one finds
\[
r \ddot r-\big|\dot F_{\perp}\big|^2=\langle\ddot F,F\rangle ,
\]
where 
\[
\dot F_{\perp} : = \dot F -  \left \langle \dot F , \frac{ F}{\big| F \big| }  \right \rangle  \frac{ F}{\big| F \big| } 
\]
and hence
\[
\big|\dot F_{\perp}\big|^2= \dot F^2 - \left \langle \dot F ,  \frac{ F}{\big| F \big| }  \right \rangle^2.
\]
This mean that the differential equation of \eqref{eq:DiffInequation} can be recast as
\begin{align}
r \ddot r-\big|\dot F_{\perp}\big|^2 \ge -\frac{1}{R^2}r^2. 
\label{eq:TowardsDiffInequality}
\end{align}
Observe that for the function $\bar F$ defined in Section \ref{section:onTheSphere} one has that $\dot{\bar F}_{\perp} =0$.

Because, $\big|\dot F_{\perp}\big|^2 \geq 0$, the equality \eqref{eq:TowardsDiffInequality} yields the following inequality
\[\ddot {r} +\frac{1}{R^2}r \ge 0,\] 
as long as $r \neq 0$, while the same inequality is trivially satisfied if $r$ is identially zero on some open subset, which means that the differential inequality is satisfied almost everywhere. 
In other words \eqref{eq:DiffInequation} yields, 
\begin{equation}\label{eq.TheDifferentialInequation}
\ddot r \;+\;\frac{1}{R^2}r \;\ge\;0, \quad\text{with} \quad r(0) = 0 \,.
\end{equation}

\DifferentialInequalityONE*

The result follows from Sturm-Liouville theory, see e.g. \cite[Chapter 8]{coddington1955theory} for a general introduction. We follow the exposition of \cite[Section 3.2]{Berger}, whose formulation is the closest to what we need, with the addition of small observation which makes sure the argument goes through if $r$ is only $C^{1,1}$ instead of $C^2$. This observation is based on the fundamental theorem of calculus for absolutely continuous functions, which can be formulated as:
\begin{theorem}[Adapted from Theorem 6.4.2 in \cite{heil2019introduction}]\label{theorem:FundamentalThmCalculusACFunctions}
If $f:[a,b] \rightarrow \R$, then the following two statements are equivalent.
\begin{itemize}
\item[(a)] $f$ is absolutely continuous.
\item[(b)] $f$ is differentiable almost everywhere on $[a,b]$, $f'\in L^1[a,b]$, and
\[
f(x) - f(a) = \int_a^x f'(t) dt.
\]
\end{itemize}
\end{theorem}
We recall that Lipschitz functions are, in particular, absolutely continuous. 
We proceed with the proof of Lemma \ref{lemma.DifferentialInequality_1}.
\begin{proof} 
We write $f$ for the solution to 
\begin{align*}
\ddot{f}  +\frac{1}{R^2}f  &= 0 &  \textrm{ with }  f(0)=0 \textrm{ and } \dot{f}(0) = \dot{r}_0 ,
\end{align*}
that is $f(t) = R\dot r(0)\sin\frac{t}{R}$.  
We examine the ratio $r/f$. Its derivative is finite and has the same sign as $N_o (t) =\dot{r} f - \dot{f} r$, as long as $t < R\pi$. The function $N_o$ is Lipschitz, because $r$ is assumed to be $C^{1,1}$.  
Calculating the derivative of $N_o$ and combining this with the differential (in)equalities yields that almost everywhere
\begin{align*} 
\ddot{r} f - \ddot{f} r\geq -\frac{1}{R^2} rf +\frac{1}{R^2} rf = 0 .
\end{align*} 
In other words the derivative is non-negative, which means that by Theorem \ref{theorem:FundamentalThmCalculusACFunctions} the function $N_o (t)$ is monotone. Moreover, the initial conditions give that $N_o(0)=0$, so that $N_o$ is always positive. This implies that the ratio $r/f$ doesn't decrease, while it is $1$ at $t=0$ by l'H{\^o}pital's rule, so that $r/f\geq 1$, or $r\geq f$, as long as $t < R\pi$. 
\end{proof}

\subsubsection{Euclidean width of an isosceles wedge}
From the previous sections and Lemma \ref{lemma.DifferentialInequality_1} in particular we have: 
\begin{lemma}\label{lemma.EuclideanGeodesicDivergence}
Consider  $a,b_1,b_2 \in \Set$ such that $d_{\Set}(a,b_i)< \pi R$, for $i=1,2$ and  denote $L:= \min(d_{\Set}(a,b_1),d_{\Set}(a,b_2))$. 
Then:
\begin{equation}\label{eq.EuclideanGeodesicDivergence}
\forall \rho \in [0,L], \; \| \gamma_{a,b_2}(\rho ) - \gamma_{a,b_1} (\rho ) \|  \geq  2 R\sin \frac{\angle \dot \gamma_{a,b_1}(0), \dot \gamma_{a,b_2}(0)}{2}   \sin\frac{\rho }{R}.
\end{equation}
\end{lemma}
The example in Section \ref{section:onTheSphere} shows that \eqref{eq.EuclideanGeodesicDivergence} is tight.
We stress that the notation $\rho$ is chosen in light of the (future) application in \eqref{eq:DifLengthSphere}. 
\subsection{Inverse exponential map and Gauss Lemma}

In his section, we prove the main theorem:
\MainCAT*

\subsubsection{Inverse exponential map and  $\ulog_a$}
We denote by $B_{\Set}^{\circ} (a,  r)$ the open geodesic ball in $\Set$ centered at $a$ with radius $r$.
Given a  point $a\in\Set$, we define the map
 \[
 \ulog_a : B_{\Set}^{\circ} (a,  \pi R) \setminus \{a\}  \rightarrow \Sphere^{d-1} \cap \Tan_{\Set}(a)
 \]
by
\[
\ulog_a (x) := \dot \gamma_{a,x}(0).
\]

\begin{lemma}\label{lemma.TauIsLIpschitz}
The map $\ulog_a$ is locally Lipschitz in $B_{\Set}^{\circ} (a,  \pi R) \setminus \{a\} $.
\end{lemma}
\begin{proof}
Take $b_i \in B_{\Set}^{\circ} (a,  \pi R) \setminus \{a\} $, for $i=1,2$, such that $d_{\Set} (b_1,b_2) < \pi R$.
We introduce the abbreviations $\gamma_i := \gamma_{a, b_i}$ and $s_i := d_{\Set} (a,b_i)$, {where the latter notation is chosen because we use $s_i$ as the parameter of the arc length parametrized geodesic. }

By triangle inequality, we get that $|s_2 - s_1|  \leq d_{\Set} (b_1,b_2)$. {We assume without loss of generality that $s_2\geq s_1$. With this assumption the} triangle inequality also yields
\[
d_{\Set} (\gamma_2 (s_1 ),  \gamma_1 (s_1 )) \leq d_{\Set} (b_1,b_2) + |s_2-s_1| \leq 2 d_{\Set} (b_1,b_2).  
\]
The following (non-tight) bound is a direct consequence of \eqref{eq.EuclideanGeodesicDivergence} 
\begin{align*}
d_{\Sphere^{d-1}} ( \ulog_a(b_1), \ulog_a(b_2)) &= \angle \dot \gamma_1 (0), \dot \gamma_2 (0) \\
&\leq 2 \arcsin \frac{ d_{\Set} (\gamma_2 (s_1 ),  \gamma_1 (s_1 )) }{2 R\sin\frac{s_1}{R}} \\
&\leq 2 \arcsin \frac{ d_{\Set} (b_1,b_2)}{R\sin\frac{s_1}{R}} , 
\end{align*}
{which gives a local Lipschitz bound, because $\arcsin(x)$ is locally Lipschitz as long as $|x|<1$.}
\end{proof}

\begin{remark} The map:
\[
\log: B_{\Set}^{\circ} (a,  \pi R) \rightarrow \Tan_{\Set}(a)\cap B_{\R^d}^{\circ}(0, \pi R),
\]
 defined by $\log(a) := 0$ and $\log(x) := d_{\Set}(a,x) \ulog_a(x)$  when $x\neq a$, 
is locally Lipschitz and generalizes the inverse of the Riemannian exponential map.
The $\log$ is not in general injective and the corresponding $\exp$ map is in general set valued.
\end{remark}
\subsubsection{The geodesic sphere is orthogonal to radial geodesics (Gauss Lemma)}
{

In this section we prove the following:

\GeodesicSphereOrthogonalToRadialGeodesics*
\noindent The proof consists of two steps, where we start with a weaker lemma:
\begin{restatable}{lemma}{HalfContinuityOfGeodesicsTangents}
\label{lemma.HalfContinuityOfGeodesicsTangents}
Given $0< \ell< \pi R$ and $\varepsilon>0$, 
there exists $\alpha (  \ell, \varepsilon) >0$ such that for any
$a,b,b' \in \Set$,  that satisfy $d_{\Set}(a,b)=d_{\Set}(a,b')=\ell$  
and $|b' -b| < \alpha$, one has
{
\[
 \angle \,  b'-b,  \dot \gamma_{a,b} (\ell)  \geq \frac{\pi}{2} - \varepsilon.
\]
}
\end{restatable}

The proof can be found in Appendix \ref{section.ProofOfHalfContinuityOfGeodesicsTangents}.
The idea is to take a point $c = \gamma_{a,b}(\ell -h)$ for a well chosen $h$ and then observe that, by triangle inequality,  
$h=d_{\Set}(c,b) \leq d_{\Set}(c,b')$.

\begin{proof}[Proof of Lemma \ref{lemma.GeodesicSphereOrthogonalToRadialGeodesics}]
From Corollary \ref{Cor:ContinuousGc1}, we know that there exists $\alpha_1>0$ such that
\begin{equation}\label{eq.proofOfGeodesicSphereOrthogonalToRadialGeodesics_eq1}
|b' - b| < \alpha_1 \Rightarrow \angle \dot \gamma_{a,b} (\ell),  \dot \gamma_{a,b'} (\ell) < \varepsilon/2.
\end{equation}
Applying Lemma \ref{lemma.HalfContinuityOfGeodesicsTangents} twice, once as it is formulated and once with $b$ and $b'$ interchanged,
we see that there exists some $\alpha_2>0$ so that the following two statements hold: 
\begin{align}
|b' - b| < \alpha_2 &\Rightarrow  \angle b'-b,  \dot \gamma_{a,b} (\ell)  \geq \frac{\pi}{2} - \varepsilon/2
\label{eq.proofOfGeodesicSphereOrthogonalToRadialGeodesics_eq2} \\
|b' - b| < \alpha_2 &\Rightarrow  \angle b-b',  \dot \gamma_{a,b'} (\ell)  \geq \frac{\pi}{2} - \varepsilon/2. 
\label{eq.proofOfGeodesicSphereOrthogonalToRadialGeodesics_eq3}
\end{align}
Observe that \eqref{eq.proofOfGeodesicSphereOrthogonalToRadialGeodesics_eq3} can be recast as
\begin{equation}\label{eq.proofOfGeodesicSphereOrthogonalToRadialGeodesics_eq4}
|b' - b| < \alpha_2 \Rightarrow \angle b'-b,  \dot \gamma_{a,b'} (\ell)  \leq \frac{\pi}{2} + \varepsilon/2. 
\end{equation}

Taking $\alpha = \min(\alpha_1, \alpha_2)$, first inequality in \eqref{eq.GeodesicSphereOrthogonalToRadialGeodesics} is \eqref{eq.proofOfGeodesicSphereOrthogonalToRadialGeodesics_eq2} and the second follows from the triangle inequality, 
 \eqref{eq.proofOfGeodesicSphereOrthogonalToRadialGeodesics_eq1},
and \eqref{eq.proofOfGeodesicSphereOrthogonalToRadialGeodesics_eq4}, which completes the proof of the lemma.
\end{proof}
\subsubsection{The proof of the second main theorem
}

We are now ready to proceed with the proof of the second main result of this paper.

We consider  $3$ points $a,b,c$ in $\Set$ such that
\begin{equation}\label{eq.TrianglePerimeterLessThanTwoPiR}
d_{\Set}(a,b) + d_{\Set}(b,c) + d_{\Set}(c,a) < 2 \pi R.
\end{equation}

It follows from the triangle inequality that no pair of points in $\{a,b,c\}$ lies at geodesic distance greater than or equal to $\pi R$. Using the first main result of the paper it therefore follows that they are joined by unique minimizing geodesics of length smaller than $\pi R$. We want to upper bound the length of the curve $\ulog_a \circ \gamma_{b,c}$ which is a stronger bound, than a bound on the Euclidean angle at $a$.
According to \cite[Chapter II.1, Proposition 1.7, claim (4)]{bridson2013metric}, 
upper bounding the angle at a triangle vertices for given edge lengths allows to characterize $\operatorname{CAT}(K)$ metric spaces.


Denote by $\rho : [0, d_{\Set}(b,c) ] \rightarrow [0, \pi R)$ the function defined by:\
\[
\rho(s) := d_{\Set}(a, \gamma_{b,c} (s) ).
\]
Observe that by triangle inequality, one has both $\rho(s) \leq d_{\Set}(a,b) + s$ and $\rho(s) \leq d_{\Set}(a,c) + d_{\Set}(b,c) - s$,
 so that \eqref{eq.TrianglePerimeterLessThanTwoPiR} implies
 \[
 \forall s \in [0, d_{\Set}(b,c) ], \; \rho(s) < \pi R.
\]


%
\begin{lemma}\label{lemma.BoundDerivativeInnerAngle}
For $s$ almost everywhere in $[0, d_{\Set}(b,c) ]$, one has
\begin{equation}\label{eq.BoundDerivativeInnerAngle}
\left| \frac{d}{dt}_{\mid t = s} \ulog_a \circ \gamma_{b,c} (t) \right| \leq  \frac{\sqrt{1 - \dot \rho(s)^2} }{R \sin \frac{\rho(s)}{R} } . 
\end{equation}
In particular, one has
\begin{equation}\label{eq.IntegralBoundInnerAngle}
\length( \ulog_a \circ \gamma_{b,c} ) \leq \int_0^{d_{\Set}(b,c)} \frac{\sqrt{1 - \dot \rho(s)^2} }{R \sin \frac{\rho(s)}{R} } ds. 
\end{equation}
\end{lemma}
\begin{proof}[Proof]
By Lemma \ref{lemma.TauIsLIpschitz} the map $\ulog_a$ is Lipschitz and $\gamma_{b,c}$ is $1$-Lipschitz, so that 
$\ulog_a \circ \gamma_{b,c} $ is Lipschitz and therefore, as a consequence of Rademacher's theorem, differentiable almost everywhere. 

The function $\rho$ is $1$-Lipschitz by the triangle inequality and is therefore also 
differentiable almost everywhere. 
Let $s \in (0, d_{\Set}(b,c)  )$ be such that both $ \ulog_a \circ \gamma_{b,c}$ and $\rho$ are differentiable at $s$ and assume moreover that
\[
\frac{d}{dt}_{\mid t = s} \rho(t) >0.
\]
{As a consequence of the previous assumption there exists an $\eta>0$ such that $t\in (s, s+\eta] \Rightarrow \rho(t) > \rho(s)$.}

\noindent
Let us define $\phi: [s, s+\eta] \rightarrow \Set$ as follows
\[
\phi(t) := \gamma_{a, \gamma_{b,c} (t)} (\rho(s) ). 
\]
We stress that, because the geodesics are arc length parametrized, $\phi(t) $ remains at a constant distance $\rho(s)$ from $a$, that is 
\begin{align} 
d_{\Set} (a, \phi(t)) = \rho(s)
\label{eq:equalDist}
\end{align} 
For $h\in (0, \eta)$ we consider the triangle with vertices $\gamma_{b,c}(s)$, $\gamma_{b,c}(s+h)$, $\phi(s+h)$.

 \begin{figure}[ht!]
\centering
\includegraphics[width= 0.85 \textwidth]{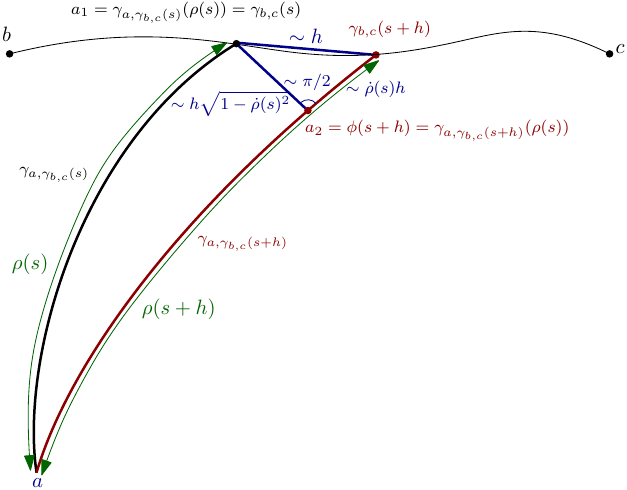}
\caption{The notation for  \eqref{eq.BoundDerivativeInnerAngle}. In the figure the $\sim$ indicates the limit as $h$ tends to zero. We stress that $\phi(s+h)$ is well defined because we assumed that $\frac{d}{dt}_{\mid t = s} \rho(t) >0$.  } 
\label{figure:LengthElement} \label{figure:Question}
\end{figure}

One has 
\begin{align}
\lim_{\substack {h\rightarrow 0 \\ h> 0} }& \frac{| \gamma_{b,c}(s+h) - \gamma_{b,c}(s) |}{h} = 1 
\tag{since geodesics are arc length parametrized} 
\\
\lim_{\substack {h\rightarrow 0 \\ h> 0} }& \frac{| \gamma_{b,c}(s+h) - \phi(s+h) |}{h} 
= \lim_{\substack {h\rightarrow 0 \\ h> 0} } \frac{\rho (s+h) - \rho(h) }{h} = \frac{d}{dt}_{\mid t = s} \rho(t)  
\tag{{see Figure \ref{figure:Question}, the first equality follows from \eqref{eq:equalDist}}} 
\\
\lim_{\substack {h\rightarrow 0 \\ h> 0} }& \angle  \gamma_{b,c}(s) - \phi(s+h), \gamma_{b,c}(s+h) - \phi(s+h) = \frac{\pi}{2}. \tag{by Lemma \ref{lemma.GeodesicSphereOrthogonalToRadialGeodesics}}
\end{align}
{These three observations combined with Pythagoras, see Figure \ref{figure:LengthElement}, yield that,} 
\begin{equation}\label{eq.DerivativePolarDIsplacement}
\lim_{\substack {h\rightarrow 0 \\ h> 0} } \frac{| \phi(s+h)  - \gamma_{b,c}(s) |}{h} = \sqrt{1 -  \left( \frac{d}{dt}_{\mid t = s} \rho(t) \right)^2 }.
\end{equation}


We now have the following
\begin{align}
&\left| \frac{d}{dt}_{\mid t = s} \ulog_a \circ \gamma_{b,c} (t) \right| 
\nonumber 
\\
&= \lim_{\substack {h\rightarrow 0 \\ h> 0} } \frac{ d_{\Sphere^{d-1}} ( \ulog_a ( \gamma_{b,c}(s+h))  , \ulog_a ( \gamma_{b,c}(s) )) }{h}  
\tag{by definition of the limit}
\\
&= \lim_{\substack {h\rightarrow 0 \\ h> 0} }\frac{ d_{\Sphere^{d-1}} (\ulog_a ( \phi(s+h)) , \ulog_a ( \gamma_{b,c}(s) )) }{h }  
\tag{by definition of $\phi$} 
\\
&\leq \lim_{\substack {h\rightarrow 0 \\ h> 0} } \frac{2}{h} \arcsin  \frac{ | \phi(s+h) -  \gamma_{b,c}(s) | }{ 2  \left( R \sin \frac{\rho(s)}{R} \right) } 
\tag{by \eqref{eq.EuclideanGeodesicDivergence}, since both geodesics have length $\rho(s)$}
\\
&= \lim_{\substack {h\rightarrow 0 \\ h> 0} }   \frac{ | \phi(s+h) -  \gamma_{b,c}(s) | }{  h \left( R \sin \frac{\rho(s)}{R} \right) } 
\tag{as the derivative of $\arcsin$ is $1$ at zero} 
\\ 
& = \frac{  \sqrt{1 -  {\dot \rho(s)}^2 } } { R \sin \frac{\rho(s)}{R} }. \tag{by \eqref{eq.DerivativePolarDIsplacement}}
\\
\label{eq:DifLengthSphere}
\end{align}
This with Lemma \ref{lemma.DifferentialInequality_1} gives us \eqref{eq.BoundDerivativeInnerAngle}
in the case where $\frac{d}{dt}_{\mid t = s} \rho(t) >0$, 
since $\rho(t)$ is differentiable, this (bound on the) right derivative calculated in \eqref{eq:DifLengthSphere} is the (bound on the) derivative. 
%
%
The result under the assumption that $\frac{d}{dt}_{\mid t = s} \rho(t) < 0$ is symmetric (with the the left derivative replacing the right derivative).  

This only leaves us with the special case, where $\frac{d}{dt}_{\mid t = s} \rho(t) =0$. The ideas underlying this case are not very different from the analysis in the case where $\frac{d}{dt}_{\mid t = s} \rho(t) >0$, but the statement is somewhat involved. 
For a small $h\rightarrow 0$ we define
\[
s_h :=  \min( \rho(s), \rho(s+h)).
\]
 and two points $a_1,a_2$ 
\begin{align*}
a_1(h) &:= \gamma_{a, \gamma_{b,c} (s)} ( s_h ),\\
a_2(h) &:= \gamma_{a, \gamma_{b,c} (s+h)} (s_h ).
\end{align*}
the points $a, a_1, a_2$ are the vertices of an isosceles hinge consisting of the two geodesics from $a$ to $a_1$, $a_2$, respectively, see Figure \ref{fig:notationDerivativeRhoZero}.

Because $\dot \rho(s) =0$, one has
\begin{equation}\label{eq:hOnHGoesToZero}
\lim_{\substack {h\rightarrow 0 \\ h> 0} }\frac{| s_h- \rho(s)|}{h} = 0.
\end{equation}
 \begin{figure}[ht!]
\centering
\includegraphics[width= 0.85 \textwidth]{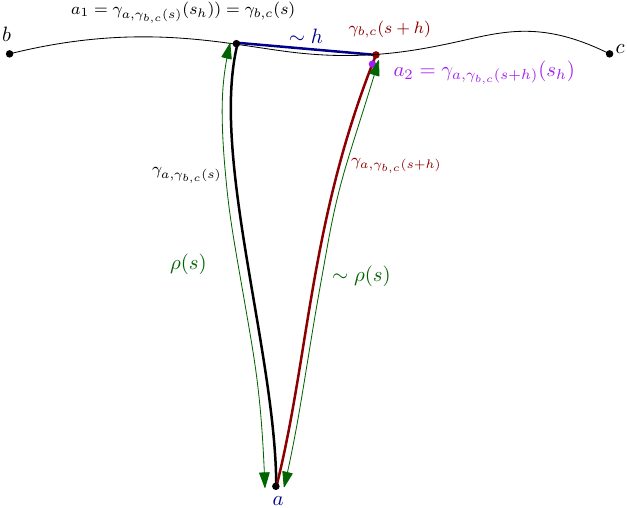}
\caption{
We have chosen to depict the case where $s_h =  \rho(s)$. } 
\label{fig:notationDerivativeRhoZero} 
\end{figure}

We see that 
\begin{align}
&\left| \frac{d}{dt}_{\mid t = s} \ulog_a \circ \gamma_{b,c} (t) \right| 
\nonumber
\\
&= \lim_{\substack {h\rightarrow 0 \\ h> 0} }\frac{ d_{\Sphere^{d-1}} ( \ulog_a( a_1(h)) , \ulog_a( a_2(h)) ) }{h }  
\tag{By definition of the derivative and $a_1$, $a_2$} 
\\
&\leq \lim_{\substack {h\rightarrow 0 \\ h> 0} } \frac{2}{h} \arcsin \frac{  |  a_2(h) -   a_1(h) | }{2 \left( R \sin \frac{{ s_h}}{R} \right) } 
\tag{by \eqref{eq.EuclideanGeodesicDivergence}, since both geodesics have length $s_h$ } \\ 
&\leq \lim_{\substack {h\rightarrow 0 \\ h> 0} }    \frac{  |  a_2(h) -   a_1(h) | }{h \left( R \sin \frac{{s_h}}{R} \right) } 
\tag{as the derivative of $\arcsin$ is $1$ at zero} 
\\
&= \lim_{\substack {h\rightarrow 0 \\ h> 0} }\frac{  |  \gamma_{b,c} (s+h) -  \gamma_{b,c} (s) | }{h \left( R \sin \frac{\rho(s)}{R} \right) } 
\tag{thanks to \eqref{eq:hOnHGoesToZero}, we have $\gamma_{b,c} (s+h)\sim a_2(h)$, $\gamma_{b,c} (s) \sim    a_1(h)$, and $\rho(s) \sim s_h$} \\
& = \frac{ 1} { R \sin \frac{\rho(s)}{R} }, \tag{since $|\dot \gamma_{b,c} (s)| = 1$} 
\end{align}
where in the explanation $\sim$ denotes equality up to leading order in $h$. 
\end{proof}

%
%
%
For the remainder of the paper the following will be our setting: 
\begin{setting} \label{set:FinalPart} 
We consider the comparison triangle, that is the $3$ points $\bar a, \bar b, \bar c$ in $R \Sphere^2$,
the geodesic $2$-sphere with radius $R$ with same edge lengths as $abc$, i.e.:
\[
d_{R \Sphere^2}(\bar a, \bar b) = d_{\Set}(a,b), \; d_{R \Sphere^2}(\bar b, \bar c) = d_{\Set}(b,c), \; d_{R \Sphere^2}(\bar c, \bar a) = d_{\Set}(c,a).
\]
\end{setting} 

\begin{lemma}\label{lemma.ExistsGammaGivenRho} 
Let $\tilde \rho : [0, d_{\Set}(b,c) ] \rightarrow (0, \pi R)$ be a $1$-Lipschitz function such that
\begin{align*} 
\tilde \rho(0) &= d_{\Set}(a,b) & & \text{and} & \tilde \rho(d_{\Set}(b,c) ) &= d_{\Set}(a,c).
\end{align*}
Then there exist a unit speed curve $\tilde \gamma:  [0, d_{\Set}(b,c) ] \rightarrow R \Sphere^2$, with $\tilde \gamma (0) = \bar b$, such that
for any $s\in [0, d_{\Set}(b,c) ]$,
\begin{align}
&\tilde \rho(s) = d_{R \Sphere^2}(\bar a, \tilde \gamma (s) ) \label{eq.DefRhoTilde}\\
&\frac{d}{dt}_{\mid t = s} \arg \ulog_{\bar a} \circ \tilde \gamma (t) \geq 0, \label{eq.ThetaIncreases}
\end{align}
where $\bar{a}$ and $\bar{b}$ are as defined in  Setting \ref{set:FinalPart} and $\arg : \Sphere^1 \rightarrow [0, 2 \pi)$ is the argument or angle with {its zero angle or argument} defined by $\arg \ulog_{\bar a} (\bar b) = 0$.
Moreover, denoting $\tilde c := \tilde \gamma (d_{\Set}(b,c))$, the end point of $\tilde \gamma$, one has
\begin{equation}\label{eq.AngleTildeOnSphereAsIntegral}
\angle \ulog_{\bar a} (\bar b), \ulog_{\bar a} (\tilde c)  = \int_0^{d_{\Set}(b,c)} \frac{\sqrt{1 - \dot {\tilde \rho}(s)^2} }{R \sin \frac{\tilde \rho(s)}{R} } ds. 
\end{equation}
\end{lemma}

\begin{proof}
We consider the spherical coordinates $(\phi,\theta) \in   (0, \pi)  \times [0, 2 \pi)$ defined for
$x\in R \Sphere^2 \setminus \{ (0,0, R  ), (0,0, - R  )\}$, where we allow ourselves to identify 
$R \Sphere^2 $ with its embedding in $\R^3$:
\[
x= 
\begin{pmatrix}
R \sin \phi(x) \cos \theta(x) \\
R \sin \phi(x) \sin \theta(x) \\
R \cos \phi(x) 
\end{pmatrix}.
\]
Without loss of generality we assume that $\bar a = (0,0,R )$ is the north pole with (extended) coordinate $\phi ( \bar a) = 0$
and $\bar b$ on { the zero or `Greenwich'} meridian, that is $\theta (\bar b) = 0$ and $\phi (\bar b) = \tilde{\rho}(0)= d_{\Set}(a,b)$,
while $\phi ( \tilde {c}) = \tilde{\rho}(d_{\Set}(b,c) ) = d_{\Set}(a,c)$.
{With these conventions we have that} 
\begin{equation}\label{eq.UlogIsTheta}
 \theta(x) = \arg \ulog_{\bar a} (x) .
\end{equation}
{Moreover, with this notation the objective \eqref{eq.DefRhoTilde} translates into showing } 
the existence of a function $\tilde \gamma$ such that $\tilde \gamma (0) = \bar a$ and,
for $s\in [0, d_{\Set}(b,c) ]$,
\begin{equation}\label{eq.PhiOfTildeGamma}
\phi (\tilde \gamma( s))  =  \frac{1}{R}\tilde \rho (s).
\end{equation}
{ We stress that,} since $\tilde \rho$ is assumed to be $1$-Lipschitz, its derivative {exists almost everywhere due to Rademacher's theorem} and belongs to $[-1,1]$.
{Additionally, we} define the normalized basis made of vector fields ``south'' and ``east'' , denoted $\frak s$ and $\frak e$, of the tangent plane for a point $x$ {in terms of the coordinates}
$(\theta, \phi) \in [0, 2 \pi) \times (0, \pi)$:
\begin{align*}
\frak s &:= \frac{1}{R} \frac{\partial x} {\partial \phi} ,&
\frak e &:= \frac{1}{R \sin \phi} \frac{\partial x}{\partial \theta} .
\end{align*}

If a such $\tilde{\gamma}$ exists, using \eqref{eq.DefRhoTilde} and \eqref{eq.PhiOfTildeGamma}, {it must almost everywhere satisfy, }  
\[
\dot {\tilde \gamma} (s) = \dot {\tilde \rho} (s) \,\frak s 
+ R \sin \frac{\tilde \rho (s) }{R} \frac{d}{dt}_{\mid t = s}  \theta ( \tilde \gamma (t)) \, \frak e
\]
Since $| \dot {\tilde \gamma} (s) | =1$,  {and by \eqref{eq.ThetaIncreases} and \eqref{eq.UlogIsTheta} $\theta$ is strictly monotone}, 
 {Pythogoras gives } 
\[
\frac{d}{dt}_{\mid t = s}  \theta ( \tilde \gamma (t)) = \frac{\sqrt {1 - \dot {\tilde \rho} (s)^2 }}
{ R \sin \frac{\tilde \rho (s) }{R} },
\]
{again, almost everywhere.
Using the fundamental theorem of calculus for absolutely continuous functions (Theorem \ref{theorem:FundamentalThmCalculusACFunctions}), again yields } 
\begin{equation}\label{eq.ThetaOfTildeGamma}
\theta ( \tilde \gamma (s)) = \int_0^s \frac{\sqrt {1 - \dot {\tilde \rho} (t)^2 }}
{ R \sin \frac{\tilde \rho (t) }{R} } dt.
\end{equation}
{
We can now define the curve $\tilde \gamma$ as having the coordinates $(\phi,\theta)$ given in \eqref{eq.PhiOfTildeGamma} and  \eqref{eq.ThetaOfTildeGamma}. 
With this choice \eqref{eq.AngleTildeOnSphereAsIntegral} follows from \eqref{eq.UlogIsTheta} and \eqref{eq.ThetaOfTildeGamma} and the lemma is proven.}
\end{proof}

}

Consider $\bar \rho : (0, d_{\Set}(b,c) ] \rightarrow [0, \pi R)$ defined by:
\begin{align} 
\bar \rho(s) := d_{R \Sphere^2} (\bar a, \bar \gamma_{\bar b, \bar c} (s) )
\label{eq:defrhoB}
\end{align} 
where $\bar \gamma_{\bar b, \bar c}$ is the minimizing geodesic, on $R \Sphere^2$, from $\bar b$ to $\bar c$.

\begin{lemma}\label{lemma.BoundAngleOnSphere} 
The function $\bar \rho$ {of \eqref{eq:defrhoB}} maximizes 
the right hand side of~\eqref{eq.IntegralBoundInnerAngle}, that is 
\[\int_0^{d_{\Set}(b,c)} \frac{\sqrt{1 - \dot \rho(s)^2} }{R \sin \frac{\rho(s)}{R} } ds, 
\]	
among all {1-}Lipschitz functions $\tilde \rho : [0, d_{\Set}(b,c) ] \rightarrow   (0, \pi R)$,
under the constraints
\begin{equation}\label{eq.BoundaryConditionsOnRho}
\tilde \rho(0) = d_{\Set}(a,b) \quad \text{and} \quad \tilde \rho(d_{\Set}(b,c) ) = d_{\Set}(a,c).
\end{equation}
{ 
Moreover, the angle $\angle \ulog_{\bar a} (\bar b), \ulog_{\bar a} (\bar c)$ defined by the curve $\tilde{\gamma}$ in Lemma \ref{lemma.ExistsGammaGivenRho} with $\tilde{\rho} = \bar{\rho}$ is equal to this maximum. 
}
\end{lemma}
\begin{proof}
%

%

{
As asserted by Lemma  \ref{lemma.ExistsGammaGivenRho}, for any such function $\tilde \rho$ satisfying \eqref{eq.BoundaryConditionsOnRho} there exists a curve $\tilde \gamma$ with length $d_{\Set}(b,c)$
connecting $\bar b$ and $\tilde c$ and such that 
the right hand side of \eqref{eq.IntegralBoundInnerAngle} is equal to the angle 
$\angle \ulog_{\bar a} (\bar b), \ulog_{\bar a} (\tilde c)$. Since the length of $\tilde \gamma$ is $d_{\Set}(b,c)$, the distance 
$d_{R \Sphere^2}(\bar b, \tilde c)$ is upper bounded by $d_{\Set}(b,c)$ and this bound is attained when $\tilde \gamma$
is a geodesic.
 
%

Maximizing the angle at $\bar a$, 
that is $\angle \ulog_{\bar{a}}(\bar{b}),\ulog_{\bar{a}}(\tilde c)$,
is equivalent to maximizing the distance between $\bar b$ and $\tilde c$ which arises when 
$d_{R \Sphere^2}(\bar b, \tilde c) = d_{\Set}(b,c)$, that is when $\tilde \gamma$ is a geodesic.
In this extremal case, the spherical triangle $(\bar a, \bar b, \tilde c)$ is isometric to  $(\bar a, \bar b, \bar c)$.

The right hand side of \eqref{eq.IntegralBoundInnerAngle} for some $1$-Lipschitz function $\rho$ 
satisfying the boundary constraints \eqref{eq.BoundaryConditionsOnRho} is therefore upper bounded by 
the corresponding integral when $\rho$ is the $\bar \rho$ {of \eqref{eq:defrhoB}.}  

Thanks to \eqref{eq.AngleTildeOnSphereAsIntegral}, this upper bound is the angle $\angle \ulog_{\bar a} (\bar b), \ulog_{\bar a} (\bar c)$.
}
\end{proof}
%
%

\paragraph{Main theorem}The main theorem follows from Lemmas \ref{lemma.BoundDerivativeInnerAngle} and \ref{lemma.BoundAngleOnSphere}:
\begin{proof}[Proof of the main theorem]

Using the characterization \cite[Part II, Proposition 1.7, claim (4)]{bridson2013metric} of CAT(K) spaces, recalled in Section \ref{sec:notation},
it suffices to prove that the angles at the vertices of triangle $abc$ with perimeter less than $2 \pi R$
are not greater than the angle of a spherical triangle with same edge lengths on a $2$-sphere with  radius $R$. 
The angle $\angle \dot \gamma_{a,b}(0),  \dot \gamma_{a,c}(0)$ is the (geodesic) distance on the sphere between the end points of the curve 
$\ulog_a \circ \gamma_{b,c} $ on $\Sphere^{d-1}$. The length of this geodesic therefore upper bounds 
{this angle.

Observe that, by the triangle inequality, the function $\rho$ appearing in \eqref{eq.IntegralBoundInnerAngle} is
$1$-Lipschitz, satisfies \eqref{eq.BoundaryConditionsOnRho}, and has values inside the open interval $(0, \pi R)$.
It follows, that, by \eqref{eq.IntegralBoundInnerAngle}, angle $\angle \dot \gamma_{a,b}(0),  \dot \gamma_{a,c}(0)$ is}
itself upper bounded by the maximal possible value of
\[
 \int_0^{d_{\Set}(b,c)} \frac{\sqrt{1 - \dot {\tilde \rho}(s)^2} }{R \sin \frac{{\tilde \rho}(s)}{R} } ds,
\]
among all {1-}Lipschitz functions ${\tilde \rho} : [0, d_{\Set}(b,c) ] \rightarrow  {(} 0, \pi R)$,
under constraints \eqref{eq.BoundaryConditionsOnRho}. 
In turn, Lemma \ref{lemma.BoundAngleOnSphere}, says that this maximal value is precisely the angle at $\bar a$ of the comparison triangle ${\bar a}{\bar b}{\bar c}$.
 Of course, swapping vertex $a$ with $b$ or $c$ gives the same upper bounds on the corresponding angles.
 \end{proof}
\begin{remark}
The proof in fact prove a slightly stronger result than mere ``$\operatorname{CAT}$ness'', namely
 the upper bound on the length of $\ulog_a \circ \gamma_{b,c} $ on $\Sphere^{d-1}$.
\end{remark}


\section*{Acknowledgements} 
We thank Jean-Daniel Boissonnat for discussion and Eddie Aamari for his insight in possible applications. We are indebted to the reviewers (SoCG26) for many suggestion and in particular noting that Corollary \ref{Cor:CorollaryOfCorollary} immediately follows from the results in our submission.

\phantomsection
\addcontentsline{toc}{section}{Bibliography}
\bibliographystyle{alpha}
\bibliography{geomrefs}

\appendix

\section{Proofs of technical lemmas}

\subsection{Proof of Lemma \ref{lemma.SecondDerivaticeIInNor}} 
\label{section.ProofLemmaSecondDerivaticeIInNor}


\SecondDerivaticeIInNor*

\begin{proof}[Proof of Lemma \ref{lemma.SecondDerivaticeIInNor}]
We first consider the case where $\gamma_{a,b}$ is twice differentiable {at parameter $s_0$} and prove \eqref{eq.SecondDerivativeInNor}.

We consider the arc length parametrized geodesic $\gamma : [0, \ell]  \rightarrow \Set$, which is by Corollary \ref{cor:TwiceDiffAlmostAnywhere} a $C^{1,1}$ curve with reach at least $R$. To derive a contradiction we assume that, for some parameter $s_0 \in (0, \ell )$, the second order derivative $\ddot \gamma(s_0)$ exists and one has  $\ddot{\gamma}(s_0) \notin \Nor_{\Set}( \gamma(s_0))$.
We will build an infinitesimal variation of $\gamma$ in a neighborhood of $\gamma(s_0)$ that shortens $\gamma$. 
As a result $\gamma$ cannot be a minimizing geodesic, even locally, that is it is not a local minimizer of the length.

Since $0 \in \Nor_{\Set}( \gamma(s_0) )$, 
our assumption gives us that $\ddot{\gamma}(s_0)\neq 0$.

We introduce the following notation 
\[
x := \gamma(s_0), \quad u := \dot \gamma(s_0), \quad v := \frac {\ddot{\gamma}(s_0)}{| \ddot{\gamma}(s_0) |}
\]
and write $\kappa:= | \ddot{\gamma}(s_0) | >0 $ for the (scalar) curvature, see e.g. \cite{ONeill},  
so that 
\[
\ddot{\gamma}(s_0) = \kappa v. 
\]
Because $\Nor_{\Set} (x)$ is the dual cone of $\Tan_{\Set}(x)$, there exists $\tau_1 \in  \Tan_{\Set}(x)$, 
such that  
{
\begin{align} 
\label{eq:deftau1} \langle \tau_1, \ddot{\gamma}(s_0) \rangle > 0.
\end{align} 
} 


We define
\[
\tau := \frac{ \tau_1 - \langle u, \tau_1 \rangle  u }{ | \tau_1 - \langle u, \tau_1 \rangle  u | } . 
\]
Since $s_0 \in (0,\ell)$, we have that $- u , u  \in \Tan_{\Set}( x)$. Moreover, $\Tan_{\Set}( x )$ is convex and thus $\tau \in\Tan_{\Set}( x )$, because $\pm u, \tau_1 \in \Tan_{\Set}( x )$. By construction we have that $|\tau | = 1$ and $\langle u, \tau \rangle = 0$.   
Since $|\dot{\gamma}|^2 = 1$ we see that $\langle u, v \rangle = 0$, which if combined with \eqref{eq:deftau1} gives
\[ \langle \tau, v \rangle,  \langle \tau, \ddot{\gamma}(s_0) \rangle >  0.\] 
%
%
%

The curve $\gamma$ can be parametrized by its projection on the line tangent to $\gamma$ at $x$, we call this parametrization 
$\hat{\gamma}$, 
that is,
\[
\hat\gamma(t) := \gamma (s(t)) = x + t u + \frac{1}{2} \kappa t^2 v + \delta_{\tau}(t)\tau +  w_{\delta} (t) ,
\]
where $ \delta_{\tau}(0) = \frac{d}{dt} \delta_{\tau}(0) = \frac{d^2}{dt^2} \delta_{\tau}(0) =  0$ and  $w_\delta(t)$ is orthogonal to $u$ and $\tau$ and $|w_{\delta}(0)|=|\dot{w}_{\delta}(0)|= |\ddot{w}_{\delta}(0)|=0$. 

Let us denote
\[
 \beta := \frac{\pi}{2} - \angle \,  \tau , \ddot{\gamma}_{a,b}(s) >0.
\]
We further define $w_0$ to be the unit vector 
\[ w_0:= \frac{v- \langle v, \tau\rangle \tau}{ |v- \langle v, \tau\rangle \tau|}, 
\]
which is orthogonal to $\tau$. 
With these definitions we have that 
\[
v = \cos \beta w_0 + \sin \beta \tau, 
\]
and we can recast $\hat{\gamma}$ in the form
\[
\hat\gamma(t) := \gamma (s(t)) = x + t u + \left (\frac{1}{2} \kappa t^2 \sin \beta  + \delta_{\tau}(t) \right)\tau + \frac{1}{2} \kappa t^2 \cos \beta w_0 +  w_{\delta} (t) ,
\]

Take an angle $\theta$ such that
\begin{equation}\label{eq.boundTheta}
0< \sin \theta< \frac{R \kappa \sin \beta}{20} .
\end{equation}
We'll now exploit the semi-continuity of the tangent and normal cones established in \cite{FullVerionc11}. 
More precisely, by \cite[equation (11)]{FullVerionc11}, there exists $\alpha_1 >0$ with $\alpha_1 < R/10$ such that,
\begin{equation}\label{eq.AngleToTangentConeLetThanTheta}
\forall y \in B(x , \alpha_1) \cap \Set,  \,    \angle \, \tau, \Tan_{\Set}( x) < \theta.
\end{equation}

There is an $\alpha$ satisfying $0<\alpha < \alpha_1/2$ such that for all $ t \in [-\alpha, \alpha]$ we have that 
\begin{align}\label{eq.BoundOnDeltaDerivatives}
\left| \frac{d}{dt} \delta_{\tau}(t) \right| ,  \left| \dot{w}_\delta(t) \right|   < \frac{1}{10}\kappa \alpha \sin \beta 
\end{align}
We note that 
\[
\frac{d}{dt} \hat\gamma(t) = u + ( \kappa t \sin \beta +  \dot\delta_{\tau}(t) ) \tau +  \kappa t \cos \beta w_0+  \dot{w}_{\delta}(t) ,
\]
where, as mentioned, all vectors are orthogonal with the exception of $w_0$ and $\dot{w}_{\delta}(t)$.
Hence we can express the length of $\gamma([s(-\alpha), s(\alpha)] )$ as 

\begin{align*}
\length \, &\hat\gamma ([-\alpha, \alpha ] ) 
\\ &= s(\alpha) - s(-\alpha) 
\\
&= \int_{-\alpha}^{\alpha} \sqrt{ 1 + ( \kappa t \sin \beta +  \dot\delta_{\tau}(t) )^2 + ( \kappa t \cos \beta)^2 + | \dot{w}_\delta(t) |^2 +2 \kappa t \cos \beta \langle \tau , \dot{w}_\delta(t) \rangle }\: dt
\end{align*} 
For $\lambda \in [0, \sin \beta]$ and $t\in [-\alpha, \alpha]$ we consider the following perturbation of $\hat\gamma$, denoted by  $\hat\gamma_{\lambda}$, 
\[
\hat\gamma_{\lambda} (s(t)) = \hat\gamma(t) + \lambda \left( \frac{1}{2} \kappa ( \alpha^2 - t^2) \right) \tau.
\]
We stress that by construction $\hat\gamma_0 = \hat\gamma$, $\hat\gamma_{\lambda} (-\alpha )  = \hat\gamma(-\alpha)$, and $\hat\gamma_{\lambda} (\alpha)  = \hat\gamma(\alpha)$. As in the first part of this paper this variation is not (necessarily) a curve in $\Set$, but its closest point projection is, which we will come back to later. 

One has
\begin{align*}
&\length \hat\gamma_{\lambda}  ([-\alpha, \alpha] ) \\
&\quad = \int_{-\alpha}^{\alpha} \sqrt{ 1 + ( \kappa t \sin \beta +  \dot\delta_{\tau}(t) - \lambda  \kappa t )^2 + ( \kappa t \cos \beta)^2 + ( \dot{w}_\delta(t) )^2 +2 \kappa t \cos \beta \langle \tau , \dot{w}_\delta(t) \rangle } \: dt
\end{align*} 
and as a result we also see that
\begin{align}
&{\frac{d}{d\lambda}}_{\mid \lambda = 0} \length \hat\gamma_{\lambda}  ([-\alpha, \alpha] ) 
\nonumber
\\
&\quad = \int_{-\alpha}^{\alpha} \frac{- ( \kappa t \sin \beta +  \dot\delta_{\tau}(t) )  \kappa t}
{\sqrt{ 1 + ( \kappa t \sin \beta +  \dot\delta_{\tau}(t) )^2 + ( \kappa t \cos \beta)^2 + ( \dot{w}_\delta(t) )^2 +2 \kappa t \cos \beta \langle \tau , \dot{w}_\delta(t) \rangle } }\: dt
\label{eq:LengthInt}
\end{align} 
Since $\alpha  < R/10$ one has that $\kappa \alpha < 1/10$ which, if combined with \eqref{eq.BoundOnDeltaDerivatives}, gives that 
\begin{align}
\kappa t \sin \beta &< 1/10
\nonumber
\\
 \dot\delta_{\tau}(t) &< 1/10 \kappa \alpha \sin \beta < 1/100
\nonumber
\\
\kappa t \cos \beta &< 1/10
\nonumber
\\
\left| \dot{w}_\delta(t) \right|   &< \frac{1}{10}\kappa \alpha \sin \beta< 1/100
\nonumber
\\
2 \kappa t \cos \beta \langle \tau , \dot{w}_\delta(t) \rangle  &\leq  2 \kappa t \cos \beta | \dot{w}_\delta(t)| < 2/1000 ,
\tag{because $|\tau|=1$ by construction}  
\end{align}  
so that the denominator of \eqref{eq:LengthInt} is bounded as follows
\[
\forall t \in [-\alpha, \alpha], \: \sqrt{ 1 + ( \kappa t \sin \beta +  \dot\delta_{\tau}(t) )^2 + ( \kappa t \cos \beta)^2 + ( \dot{w}_\delta(t) )^2 +2 \kappa t \cos \beta \langle \tau , \dot{w}_\delta(t) \rangle }  < 2.
\]
This means that we have the following bound on \eqref{eq:LengthInt}
\[
{\frac{d}{d\lambda}}_{\mid \lambda = 0} \length \hat\gamma_{\lambda}  ([-\alpha, \alpha] ) 
< \int_{-\alpha}^{\alpha}  -\frac{1}{2} ( \kappa t \sin \beta +  \dot\delta_{\tau}(t) )  \kappa t\: dt
\]
Using \eqref{eq.BoundOnDeltaDerivatives} again, we see that 
\begin{align*}  \left |\int_{-\alpha}^{\alpha} \frac{1}{2} \dot\delta_{\tau}(t)   \kappa t\: dt \right | 
&\leq \int_{-\alpha}^{\alpha} \frac{1}{2} \left |  \dot\delta_{\tau}(t)   \kappa t\right | dt 
\\
&\leq  \int_{-\alpha}^{\alpha} \frac{1}{2} \left( \frac{1}{10} \sin (\beta) \kappa \alpha \right)  \kappa |t|  dt  
\\
&= \frac{1}{20}  \sin (\beta) \kappa^2 \alpha^3. 
\end{align*} 
We therefore find that 
\begin{equation}\label{eq.BoundOnLengthDecreasing}
{\frac{d}{d\lambda}}_{\mid \lambda = 0} \length \hat\gamma_{\lambda}  ([-\alpha, \alpha] ) 
< - \frac{1}{4} \int_{-\alpha}^{\alpha} \kappa^2 t^2 \sin \beta dt 
= - \frac{1}{6}  \kappa^2 \alpha^3 \sin \beta .
\end{equation}
From this, in turn, it follows that, for some $\eta_1>0$, one has
\begin{align}
&\forall \lambda \in [0, \eta_1], \nonumber \\
&\quad  \frac{\length \hat \gamma_{\lambda}  ([-\alpha, \alpha] ) } { \length \hat \gamma  ([-\alpha, \alpha] ) }
 \leq  \frac{ \length \hat \gamma  ([-\alpha, \alpha ] )  - \frac{1}{8}  \lambda \kappa^2 \alpha^3 \sin \beta } {   \length \hat \gamma  ([-\alpha, \alpha ] )  }\nonumber \\
 &\quad  \leq \frac{ 2 \alpha  - \frac{1}{8}  \lambda \kappa^2 \alpha^3 \sin \beta } {   2 \alpha } = 1 - \frac{1}{16}  \lambda \kappa^2 \alpha^2 \sin \beta. \label{eq.LowerBoundLengthShorterCurve}
\end{align}

We want to prove that, for $\lambda>0$ small enough, $\pi_{\Set} \circ \hat \gamma_{\lambda}  ([-\alpha, \alpha ] )$ is strictly shorter than $\hat \gamma  ([-\alpha, \alpha ] )$ while preserving the endpoints.
For this we first need that the variation stays close to $\Set$, which is formally handled in the following claim: 
\begin{claim2}\label{claim.distToSet} Let $\hat \gamma_{\lambda} (t)$ be as defined above. We have that: 
\begin{align*}
&\forall \varepsilon >0, \, \exists \, \eta>0 \quad \text{such that,}\\
&\quad     \lambda \in [0, \eta] \Rightarrow \max_{t \in [-\alpha, \alpha] }  d_{\R^d} ( \hat \gamma_{\lambda} (t), \Set)  <   \frac{1}{2}  \lambda (\varepsilon + \sin \theta )\kappa \alpha^2 ,
\end{align*} 
for sufficiently small $\alpha$.
\end{claim2}
\begin{proof}[Proof of Claim \ref{claim.distToSet}]
Let $\varepsilon >0$. 
One can clearly assume without loss of generality that
\begin{equation}\label{EpsilonLetThanOne}
\varepsilon \leq 1{/ 2 } .
\end{equation}

By \eqref{eq.AngleToTangentConeLetThanTheta} and the definition of tangent cone, for any $t\in [-\alpha, \alpha]$, there exists $x(t) \in \Set$ such that 
\begin{align} \label{eq.DefXOfT}
0< |x(t)- \hat\gamma(t) | &< \frac{\varepsilon  R}{10} && \textrm{ and} 
&\angle \tau, x(t) - \hat\gamma(t) < \theta + \varepsilon /2.
\end{align}
Recall that $\hat \gamma(t') \in \Set$ and $\hat \gamma_{\lambda} (t') = \hat \gamma(t') +  \lambda \left( \frac{1}{2} \kappa ( \alpha^2 - t'^2) \right) \tau$.
We define
 \begin{align*}
 \eta(t) &:= \frac{     |x(t)- \hat\gamma(t) |}{ \kappa \alpha^2} \\
 h(t)&:= |x(t)- \hat\gamma(t) | \varepsilon /4.
 \end{align*}


We'll now show that for any $t'\in (t- h(t),  t + h(t) )$, the minimizing geodesic between $\hat \gamma(t')$ and $x(t)$ is
close enough to the segment $[\hat \gamma(t'), x(t)]$, c.f. Figure \ref{figure:DistClaim7}.
To this end we define
\[
\tau' = \frac{ x(t) - \hat \gamma(t')}{ |x(t) - \hat \gamma(t')| }.
\]
By \eqref{eq.DefXOfT}, one has that,
\begin{equation}\label{FDefThetaPrime}
\theta' := \angle \tau, \tau' < \theta + \varepsilon /2.
\end{equation}

 \begin{figure}[ht!]
\centering
\includegraphics[width= 0.9\textwidth]{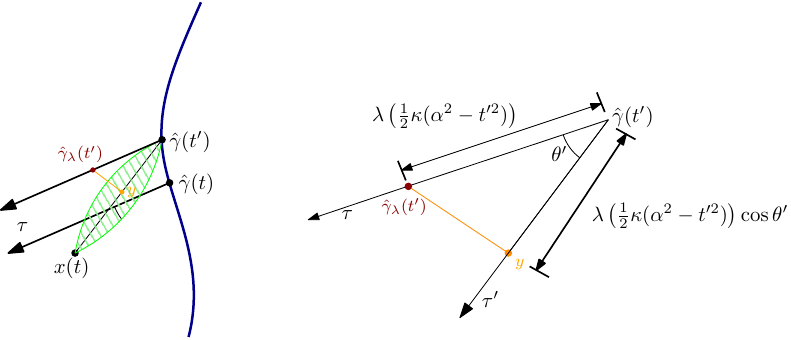}
\caption{The notation and setting of Claim \ref{claim.distToSet}.} 
\label{figure:DistClaim7}
\end{figure}

Denote by $y$ the point on the segment $[\hat \gamma (t'), x(t)]$ closest to $\hat \gamma_{\lambda} (t')$, so that, as we see in Figure~ \ref{figure:DistClaim7},  
\[
y = \hat \gamma (t') +  \lambda \left( \frac{1}{2} \kappa ( \alpha^2 - t'^2) \right) \cos \theta'  \tau'
\]
 and therefore
\begin{align}\label{eq.DistToSegmentGammaX}
| \hat \gamma_{\lambda} (t') - y | &=  \lambda \left( \frac{1}{2} \kappa ( \alpha^2 - t'^2) \right) \sin \theta' \leq \left( \frac{1}{2} \kappa ( \alpha^2 - t'^2) \right) \sin ( \theta + \varepsilon /2),
\end{align}
where we used \eqref{FDefThetaPrime} for the last inequality. 
We stress that here we used the definition of $\eta$. In fact, the definition of $\eta$ ensures that $|\hat \gamma_{\lambda} (t')-\hat \gamma (t')| \leq |\hat \gamma (t') - x(t)|$, so that the projection of $y$ onto the segment $[\hat \gamma (t'), x(t)]$ lies in the interior, and the calculations above apply. Indeed, we see that
\begin{align}
|\hat \gamma_{\lambda} (t')-\hat \gamma (t')| \leq  \lambda \left( \frac{1}{2} \kappa ( \alpha^2 - t'^2) \right) \leq  \left( \frac{1}{2} \kappa ( \alpha^2 - t'^2) \right) \eta \leq \frac{1}{2} |x(t)- \hat\gamma(t) |,
\nonumber
\end{align}  

while 
\begin{align} 
|\hat \gamma (t') - x(t)| \geq | x(t) - \hat{\gamma} (t) |  - | \hat{\gamma} (t)- \hat{\gamma} (t')| 
\label{eq:ExplanationEta1} 
\end{align}

The second term is bounded in terms of $|t-t'|$, which is in turn bounded by $h(t)$, and the Lipschitz constant of $\hat{\gamma}$. The Lipschitz constant of $\hat{\gamma}$ is determined by the angle that $\gamma$ makes with the tangent line, because $\gamma$ is $C^{1,1}$ we can assume that this constant is smaller than $2$ by decreasing $\alpha$ if necessary. Because $|t-t'|$ is upper bounded by $h(t)$, {we have that} 
{
\begin{align} 
| \hat{\gamma} (t)- \hat{\gamma} (t')| \leq 2 h(t) 
\nonumber 
\end{align} 
and {\eqref{eq:ExplanationEta1} yields}
}
%
\begin{align} 
|\hat \gamma (t') - x(t)| 
\geq | x(t) - \hat{\gamma} (t) |  -    \frac{\varepsilon}{2}     | x(t) - \hat{\gamma} (t) | \geq 3/4 | x(t) - \hat{\gamma} (t) | 
\nonumber 
\end{align}

Note that, as a consequence of Lemma \ref{lemDistProjMidpoint}, if $z$  lies on the segment $[p,q]$ with $p,q \in \Su$, then the distance from $z$ do the geodesic between $p$ and $q$ on $\Su$
is upper bounded by
\[
d(z, \Su) \leq | z - p | \frac{|q-p|}{2R}.
\]
This  gives us an upper bound on the distance from $y$ to the geodesic segment in $\Su$ between $\hat \gamma (t')$ and $x(t)$, that is, 
\begin{equation}\label{FirstBoundOnDistYToS}
d(y, \Su) \leq \lambda \left( \frac{1}{2} \kappa ( \alpha^2 - t^2) \right) \cos \theta' \,   \frac{ | x(t) - \hat \gamma (t') |}{ 2R} .  
\end{equation}

By the triangle inequality we have that 
\[ 
| x(t) - \hat \gamma (t') | \leq | x(t) - \hat \gamma (t)| +  |  \hat \gamma (t) - \hat \gamma (t')|.
\] 
By \eqref{EpsilonLetThanOne} and \eqref{eq.DefXOfT} we have that the first term is bounded by $R \varepsilon/10$. 
As we have seen above 
the second term is bounded 
by $2|t-t'|\leq 2 h(t) = |x(t)- \hat\gamma(t) | \varepsilon /2 \leq \frac{\varepsilon  R}{10} \varepsilon /2$. 

This means that we find
\[
| x(t) - \hat \gamma (t') | \leq   R \varepsilon /10 + 2 R \varepsilon^2 /10<R \varepsilon /2,
\]
where we used \eqref{eq.DefXOfT} to bound $h(t)$. 
This means that \eqref{FirstBoundOnDistYToS} 
gives us
\[
d(y, \Su) \leq \lambda \left( \frac{1}{2} \kappa ( \alpha^2 - t^2) \right) \cos \theta' \varepsilon/4 \leq \lambda \left( \frac{1}{2} \kappa ( \alpha^2 - t^2) \right) \varepsilon/2 
\]
which, together with \eqref{eq.DistToSegmentGammaX} and the triangle inequality yields
\begin{align*}
 d(\hat \gamma_{\lambda} (t'), \Su) &\leq \lambda \left( \frac{1}{2} \kappa ( \alpha^2 - t^2) \right) \left( \varepsilon/2  + \sin (\theta + \varepsilon/2 ). \right) \\
 &\leq \lambda \left( \frac{1}{2} \kappa ( \alpha^2 - t^2) \right)  \left(  \varepsilon + \sin (\theta ) \right)
\tag{since $\sin(x)$ is $1$-Lipschitz} 
\\
 &\leq \lambda \left( \frac{1}{2} \kappa \alpha^2 \right) \left(  \varepsilon + \sin (\theta ) \right)
\end{align*}
We have shown that, for any $t\in [-\alpha, \alpha]$, one has:

\[
\lambda \in [0, \eta(t)] \Rightarrow \max_{t'\in (t- h(t),  t + h(t) )}
  d_{\R^d} (\hat \gamma_{\lambda} (t'), \Set)  <   \frac{1}{2}  \lambda \kappa \alpha^2 (\varepsilon + \sin \theta )
 \]
Since  $[-\alpha, \alpha] $ is compact, we can extract a finite open cover: 
\[
[-\alpha, \alpha] \subset \bigcup_{i=1,\dots, N} (t_i- h(t_i),  t_i + h(t_i) )
\]
Taking $\eta = \min _{i=1,\dots, N} \eta(t_i)$ proves the claim.
 \end{proof}

Taking $\varepsilon = \sin \theta$ in the claim gives us that, for some $\eta>0$,  one has:
\[
\lambda \in [0, \eta] \Rightarrow \max_{t \in [-\alpha, \alpha] }  d_{\R^d} ( \hat \gamma_{\lambda} (t), \Set)  <   \lambda \kappa \alpha^2 \sin \theta 
\]
using the same argument as in the first part of the paper namely that Theorem 4.8(8) of \cite{Federer} says the closest point projection $\pi_\Su$ at a point $x$ is $\frac{1}{1- d(x, \Su) /R}$-Lipschitz, we find that 
\[
\forall \lambda \in [0, \eta], \frac { \length \pi_{\Set} \circ \hat \gamma_{\lambda}  ([-\alpha, \alpha ] )} { \length \hat \gamma_{\lambda}  ([ -\alpha, \alpha] )} \leq  \frac{R}{R - \lambda \kappa \alpha^2 \sin \theta} 
=  \frac{1}{1 - \frac{1}{R} \lambda \kappa \alpha^2   \sin \theta}
\]
with\eqref{eq.LowerBoundLengthShorterCurve} and  \eqref{eq.boundTheta} we get:
\[
\forall \lambda \in [0, \eta], \frac { \length \pi_{\Set} \circ \hat \gamma_{\lambda}  ([-\alpha,  \alpha] )} 
{ \length \hat \gamma  ([-\alpha, \alpha ] )} \leq  \frac{1 - \frac{1}{16}  \lambda \kappa^2 \alpha^2 \sin \beta}{1 - \frac{1}{R} \lambda \kappa \alpha^2   \sin \theta} 
\leq 1 - \frac{1}{80}  \lambda \kappa^2 \alpha^2 \sin \beta
\]
 It follows that $\gamma$ cannot be a local minimum of the length and therefore not a geodesic.

\paragraph{Extension to Clarke's gradient.} We have proven the result, namely \eqref{eq.SecondDerivativeInNor} for any $s$ 
where $s \mapsto \gamma(s)$ is twice differentiable.
In order to extend it to \eqref{eq.ClarkeSecondDerivativeInNor} we use the definition of Clarke gradient as the convex hull of neighboring derivative whenever they exists (which is one of a number of equivalent definitions, see \cite[Theorem 2.5.1]{Clarke1990}) together with the semi-continuity of normal cones.

Write $\Omega \subset (0, d_{\Set}(a,b))$ for the set where $s \mapsto  \dot \gamma(s)$ is not differentiable.
In our specific, setting where we consider the Lipschitz function $s \mapsto \dot \gamma$, (one of) Clarke's definition(s) of the Clarke gradient \cite[Theorem 2.5.1]{Clarke1990} reads 
\begin{equation}\label{equation:ClarkeAsLimitOfDerivative}
\partial \dot \gamma (s) = \conv \left\{ \lim_{s_i \rightarrow s} \ddot \gamma (s_i) \mid \forall i\in \N,  s_i \notin \Omega \right\}
\end{equation}
where $\conv X$ is the convex hull of $X$.

The semi-continuity of the normal cone is stated in  \cite[Lemma 33]{FullVerionc11} as, 
 \begin{equation}\label{equation:SemiContinuityNormal}
\forall p \in \Set, \forall \epsilon >0, \exists \alpha >0 \textrm{ such that }   p' \in \Set \cap B(p, \alpha) \Rightarrow \Nor_{\Set}(p')  \subset \Nor_{\Set}(p)^{\angle \epsilon},
\end{equation}
where $\Nor_{\Set}(p')  \subset \Nor_{\Set}(p)^{\angle \epsilon}$ means that any vector in $\Nor_{\Set}(p')$ lies at angle at most $\epsilon$ from some vector in $\Nor_{\Set}(p)$.

Thanks to \eqref{eq.SecondDerivativeInNor}  and \eqref{equation:SemiContinuityNormal}, we see that any limit in \eqref{equation:ClarkeAsLimitOfDerivative}  
makes a zero angle with $\Nor_{\Set}(\gamma(s))$. Since $\Nor_{\Set}(\gamma(s))$ is closed and convex, we get \eqref{eq.ClarkeSecondDerivativeInNor}.
\end{proof}

\subsection{Proof of Lemma \ref{lemma.HalfContinuityOfGeodesicsTangents}}


\HalfContinuityOfGeodesicsTangents*

\begin{proof}[Proof of Lemma \ref{lemma.HalfContinuityOfGeodesicsTangents}]\label{section.ProofOfHalfContinuityOfGeodesicsTangents}
We first need to make some definitions and introduce notation, see Figure~\ref{figure:ProofLemma11} for an illustration.  
Take $h_0$ such that
\begin{align} 
0<h_0< \min \left( \ell,  R \varepsilon  , R/10 \right)
\label{eq:BoundsH0} 
\end{align}
and define
\[
\alpha := \frac{h_0^2}{4R} .
\]
Denote $\delta := |b' -b| < \alpha$ and
\[
h := 2 \sqrt{ \delta R}.
\]
So that $ \delta := \frac{h^2}{4R}$ and $h \leq h_0$.
Write $c = \gamma_{a,b}(\ell - h)$.

 \begin{figure}[ht!]
\centering
\includegraphics[width= 0.37\textwidth]{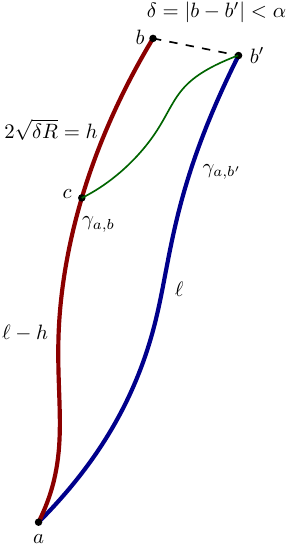}
\caption{The notation for the proof of Lemma \ref{lemma.HalfContinuityOfGeodesicsTangents}.} 
\label{figure:ProofLemma11}
\end{figure}

 Because $d_{\Set}(a, c) = \ell - h$ 
and $d_{\Set}( a, b') = \ell$, the triangle inequality gives 
\begin{equation}\label{eq.triangleInequality_0}
  h \leq d_{\Set}( c, b'), 
\end{equation}
see also Figure \ref{figure:ProofLemma11}.  
{Lemma \ref{lemma.thisLemma} now gives}
\begin{equation}\label{eq.AngleBowString}
\angle b-c, \dot \gamma_{a,b}(\ell) \leq \frac{h}{2R} \leq \frac{1}{2} \varepsilon, 
\end{equation}
{where we used the assumption \eqref{eq:BoundsH0} for the second inequality. 
Since the geodesic distance between points in $\Su$ is always greater than the Euclidean distance we see that $ | b-c| \leq h$. Furthermore, Theorem \ref{theorem:ReachEquivalentMetricDistorsion} gives }
\[
d_{\Set}( c, b')  \leq 2R \arcsin \frac{|b'-c | }{2R} \leq |b'-c| + \frac{1}{24 R^2 } |b'-c|^3 .
\]
{Combining the previous inequality with \eqref{eq.triangleInequality_0} gives} 
\begin{equation}\label{eq.inequalityTriangle_1}
|b-c| \leq |b'-c | + \frac{1}{24 R^2 } |b'-c|^3 . 
\end{equation}
Define $\theta$ to be the following angle,
\[
\theta := \angle b-c, b' -b. 
\]
The cosine rule states that 
\[
| b' -c | = \sqrt{ |b-c|^2 + |b' -b|^2- 2 |b-c| |b' -b| \cos \theta },
\]
which can be rewritten as 
\[
| b' -c | =  |b-c|  \sqrt{1 +  \frac{|b' -b|^2}{ |b-c|^2} - 
2 \frac{|b' -b|}{ |b-c|}\cos \theta }.
\]
Using that $\sqrt{1+t} \leq 1 + t/2$ one gets
\[
| b' -c | -  |b-c|  \leq   \frac{|b' -b|^2}{2  |b-c|}  -  |b' - b| \cos \theta. 
\]
Combining this with \eqref{eq.inequalityTriangle_1} gives
\[
- \frac{1}{24 R^2 } |b'-c|^3  \leq | b' -c | -  |b-c|  \leq   \frac{|b' -b|^2}{2  |b-c|}  -  |b' - b| \cos \theta,
\]
so that we get
\[
\cos \theta \leq \frac{ |b'-c|^3}{24 |b' - b| R^2 } +  \frac{|b' -b|}{2  |b-c|} .
\]
Substituting  $\frac{h^2}{4R}  \leftarrow \delta =  |b' -b|$ in to the previous expression gives 
\[
\cos \theta \leq  \frac{ |b'-c|^3}{6 h^2 R } +  \frac{h^2}{8 R  |b-c|}  .
\]
Using Theorem \ref{theorem:ReachEquivalentMetricDistorsion} again, we see that 
\begin{align*} 
|b-c| &\leq h\leq 2 R \arcsin\frac{|b-c|} {2R} \leq |b-c| + \frac{1}{24R^2} |b-c|^3 \leq |b-c| \leq \left (1+\frac{h^2}{24 R^2} \right)  |b-c|  
\\
&\leq \frac{2401}{2400} |b-c| \leq \frac{100}{99}  |b-c| ,
\end{align*} 
where for the second to last inequality we used that $h< R/10$. 
This implies that
\[
\cos \theta \leq  \frac{ (h + \delta)^3}{6 h^2 R }  +  \frac{100 h}{8 *99 R }  .
\]
We now note that $h< R/10$ and $\delta = \frac{h^2}{4R} $, which implies $\delta < h / 40$. Combining the two previous result yields 
\begin{align} 
\cos \theta \leq  \left( \frac{ 41^3}{6 *40^3 } + \frac{100 }{8 *99 } \right) \frac{h}{R} \leq 0.31 \frac{h}{R} \leq 0.31 \varepsilon
\tag{since $h < R \varepsilon$}.
\end{align} 
This in turn gives 
\[
\theta \geq \arccos 0.31 \varepsilon \geq \frac{\pi}{2} -  \frac{\varepsilon}{2}, 
\]
assuming without loss of generality that $\varepsilon$ is small enough for the  last inequality to hold.

We now arrive at
\begin{align}
\angle b'-b, \dot \gamma_{a,b}(\ell) &\geq  \angle b' -b,  b-c  \, - \, \angle b-c, \dot \gamma_{a,b}(\ell) \nonumber  \\
& \geq \theta  - \frac{\varepsilon}{2} \tag{by \eqref{eq.AngleBowString}}
\\
  &\geq \frac{\pi}{2} - \varepsilon . \label{eq.FinalAngleBound}
\end{align}

\end{proof}

\section{Alternative proof}
\begin{proof}[Alternative proof of Corollary \ref{Cor:ContinuousG}]
Assume that the geodesics are not continuous with respect to the endpoints, then there exist an $\epsilon >0$ and a pair of sequences of points $a_n$, $b_n$, where $\lim_{n \to \infty} a_n =a$ and $\lim_{n \to \infty} b_n=b$, as well as a sequence of geodesics $\gamma_n$, connecting $a_n$ and $b_n$ such that
\begin{align}  
\sup_{t\in I}  |\gamma_n(t) -\gamma(t)| \geq \epsilon,
\nonumber 
\end{align} 
where $\gamma$ is the geodesic connecting $a$ and $b$. 
We stress that here we parametrize the geodesics on the unit interval $[0,1]$, with constant norm of the tangent vector, i.e. $|\dot{\gamma}_n|= d_{\Su}(a_n , b_n)$.

We further note that thanks to the metric characterization of sets of positive reach \cite[Theorem 1]{TangentVarJournal} convergence in the Euclidean and geodesic sense are equivalent.

Again, thanks to the metric characterization of sets of positive reach \cite[Theorem 1]{TangentVarJournal} we know that geodesics are themselves sets of (at least) the same reach as the set in which they lie. We also know that manifolds of positive reach are $C^{1,1}$ where the Lipschitz constant of the derivative is bounded by $1/R$, see \cite{FullVerionc11}. This implies that the tangent vectors $\dot{\gamma}_n$ are uniformly Lipschitz, that is Lipschitz with the same Lipschitz constant for all $n$. This uniform Lipschitz constant implies by definition uniform equicontinuity of $\dot{\gamma}_n$. Because all geodesics are parametrized according to arclength, with $|\dot{\gamma}_n|= d_{\Su}(a_n , b_n)$, and $d_{\Su}(a_n , b_n)$ is bounded by assumption,  $\dot{\gamma}_n$ is also bounded. 

We now observe that by passing to a subsequence we can assume that there is one $t^* \in I$ such that 
\begin{align}  
|\gamma_n(t^*) -\gamma(t^*)| \geq \epsilon/2.
\label{DistBetweenGeodesics3}
\end{align} 
This can be seen as follows: Because $I$ is compact the supremum must be attained at some point $t^*_n$ for each $(\gamma_n -\gamma)(t)$. Again because $I$ is compact this sequence of points must have a convergent subsequence $n_m$ with limit $t^*$ because of smoothness (more precisely, the fact that $|\dot{\gamma}_n|$ is bounded) of $(\gamma_n )(t)$, we have that for $m$ sufficiently large and $|t-t^*|$ sufficiently small, we have $|\gamma_{n_m}(t) -\gamma(t)| \geq \epsilon/2 $. This implies that \eqref{DistBetweenGeodesics3} holds. 

Now we can invoke the Arzel{\`a}-Ascoli theorem to conclude that there exist a $\dot{\gamma}'$, which is the limit of a convergent subsequence of $\dot{\gamma}_n$. We write $\gamma'(t) =\int_{0}^{t}  \dot{\gamma}' \textrm{d} s$. With this notation we can formulate the following observations: 
\begin{itemize} 
\item The lengths of $\gamma_n$ converge to the length of $\gamma'$. 
\item The geodesic $\gamma_n$ has to converge to the geodesic distance between $a$ and $b$, because by the triangle inequality we have that 
\begin{align}
|\textrm{length} (\gamma) - \textrm{length} (\gamma_n) | =|d_\Su (a,b) - d_\Su (a_n,b_n)| \leq d_\Su (a,a_n) + d_\Su (b,b_n). 
\nonumber  
\end{align} 
And hence, $\gamma$ and $\gamma'$ both are geodesics, that is their lengths equal the geodesic distance between the endpoints. 
\item $|\gamma(t^*) -\gamma'(t^*)| \geq \epsilon/2$, in particular $\gamma \neq \gamma'$.
\end{itemize} 
This yields a contradiction with the uniqueness of the geodesics, that is Theorem \ref{prop:Unique}. 
\end{proof}

\end{document}